\documentclass{book}
\usepackage[utf8]{inputenc}     
\usepackage[english]{babel}     
\usepackage[a4paper, left=1in, right=1in, top=1in, bottom=1in]{geometry}
\usepackage{graphicx}           
\usepackage{amsmath,amssymb}    
\usepackage{amsthm}             
\usepackage{mathtools}          
\usepackage{enumitem}           
\usepackage{listings}           
\usepackage{todonotes}          
\usepackage{hyperref}           
\usepackage{tcolorbox}
\usepackage{tabularx}
\usepackage{array}
\usepackage{dirtytalk}

\usepackage{tikz}
\definecolor{darkgreen}{RGB}{50,205,50}
\definecolor{dred}{RGB}{165,2,36}
\definecolor{dblue}{RGB}{0,82,238}
\definecolor{dgreen}{RGB}{4,135,6}
\definecolor{dorange}{RGB}{246,106,4}

\newcolumntype{P}[1]{>{\centering\arraybackslash}p{#1}}

\usepackage{xcolor}
\usepackage{csquotes}
\usepackage{lipsum}

\usepackage[ruled,vlined, linesnumbered]{algorithm2e}

\definecolor{quotemark}{gray}{0.7}
\makeatletter
\newlength\origparskip

\newcommand{\fquote}{%
  \@ifnextchar[{\fquote@i}{\fquote@i[]}
}

\def\fquote@i[#1]{%
  \@ifnextchar[{\fquote@ii{#1}}{\fquote@ii{#1}[]}
}%

\def\fquote@ii#1[#2]{%
  \def\pqm@tempa{#1}%
  \def\pqm@tempb{#2}%
  \noindent
  \list
    {}
    {\setlength{\leftmargin}{0.3\textwidth}%
     \setlength{\rightmargin}{0.1\textwidth}%
     \setlength{\origparskip}{\parskip}}%
    \item[]%
      \begin{picture}(0,0)%
        \put(-15,-8){\makebox(0,0){\scalebox{4}{%
          \textcolor{quotemark}{\textquotedblright}}}}%
      \end{picture}%
      \begingroup
      \itshape
      \ignorespaces}%

\def\endfquote{%
  \endgroup
  \par
  \raggedleft
  \ifx\pqm@tempa\empty
  \else
    {\bfseries --- \pqm@tempa\par}%
    \setlength{\parskip}{\origparskip}%
    \ifx\pqm@tempb\empty
    \else
      (\pqm@tempb)%
    \fi
  \fi
  \par
  \endlist}
\makeatother

\def\thesistitle{On Theoretical Complexity}
\def\thesissubtitle{And Boolean Satisfiability}
\def\thesisauthorfirst{Mohamed}
\def\thesisauthorsecond{Ghanem}
\def\thesissupervisorfirst{Dr. Daoud}
\def\thesissupervisorsecond{Siniora}
\def\thesisdate{May 2021}

\title{\thesistitle}
\author{\thesisauthorfirst\space\thesisauthorsecond}
\date{\thesisdate}

\newtheorem{theorem}{Theorem}[section]
\newtheorem{corollary}{Corollary}[theorem]
\newtheorem{lemma}[theorem]{Lemma}
\newtheorem{proposition}[theorem]{Proposition}
\theoremstyle{definition}
\newtheorem{example}{Example}
\newtheorem{convention}[theorem]{Convention}

\theoremstyle{definition}
\newtheorem{definition}[theorem]{Definition}

\theoremstyle{plain} 
\newcommand{\thistheoremname}{}
\newtheorem*{genericthm}{\thistheoremname}
\newenvironment{namedthm}[1]
  {\renewcommand{\thistheoremname}{#1}%
   \begin{genericthm}}
  {\end{genericthm}}

\theoremstyle{remark}
\newtheorem*{remark}{Remark}

\begin{document}
\begin{titlepage}
	\thispagestyle{empty}
	\newcommand{\HRule}{\rule{\linewidth}{0.5mm}}
	\center
	\textsc{\Large The American University in Cairo}\\[.7cm]
	\includegraphics[width=25mm]{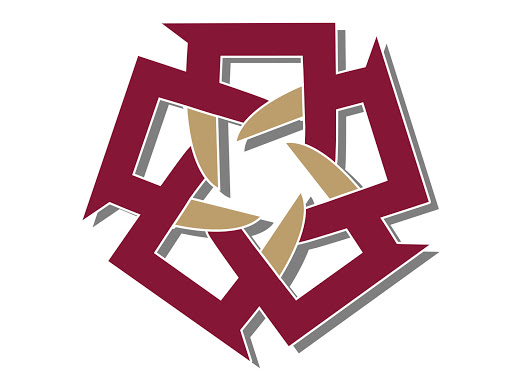}\\[.5cm]
	\textsc{Department of Mathematics and Actuarial Science}\\[0.5cm]
	
	\HRule \\[0.4cm]
	{ \huge \bfseries \thesistitle}\\[0.1cm]
	\textsc{\thesissubtitle}\\
	\HRule \\[.5cm]
	\textsc{\large Thesis BS.c. Mathematics}\\[2.5cm]
	
	\begin{minipage}{0.4\textwidth}
	\begin{flushleft} \large
	\emph{Author:}\\
	\thesisauthorfirst\space \textsc{\thesisauthorsecond}
	\end{flushleft}
	\end{minipage}
	~
	\begin{minipage}{0.4\textwidth}
	\begin{flushright} \large
	\emph{Supervisor:} \\
	\thesissupervisorfirst\space \textsc{\thesissupervisorsecond} \\[1em]
	\end{flushright}
	\end{minipage}\\[4cm]
	\vfill
	{\large \thesisdate}\\
    \clearpage
\end{titlepage}

\let\cleardoublepage=\clearpage

\newgeometry{left=1.5in, right=1.5in, top=1.5in, bottom=1.5in}

\chapter*{Abstract}
Theoretical complexity is a vital sub-field of computer science that enables us to mathematically investigate computation and answer many interesting queries about the nature of computational problems. It provides theoretical tools to assess time and space requirements of computations along with assessing the difficultly of problems - classifying them accordingly. It also garners at its core one of the most important problems in mathematics, namely, the \textbf{P vs. NP} millennium problem. In essence, this problem asks whether solution and verification reside on two different levels of difficulty.
In this thesis, we introduce some of the most central concepts in the Theory of Computing, giving an overview of how computation can be abstracted using Turing machines. Further, we introduce the two most famous problem complexity classes \textbf{P} and \textbf{NP} along with the relationship between them. In addition, we explicate the concept of problem reduction and how it is an essential tool for making hardness comparisons between different problems. Later, we present the problem of Boolean Satisfiability (SAT) which lies at the center of NP-complete problems. We then explore some of its tractable as well as intractable variants such as Horn-SAT and 3-SAT, respectively. Last but not least, we establish polynomial-time reductions from 3-SAT to some of the famous NP-complete graph problems, namely, Clique Finding, Hamiltonian Cycle Finding, and 3-Coloring.

\restoregeometry
\newgeometry{left=1.5in, right=1.5in, top=1.5in, bottom=1.5in}

\chapter*{Acknowledgement}

I would like to specially thank Professor Dauod Siniora for his faith in me, agreeing to personally supervise my thesis on this topic. I am also grateful for his insightful remarks and critique that guided me to refine this paper. I would also like to thank my friend Alaa Anani for her suggestion for me to do an undergraduate thesis in mathematics.

\restoregeometry

\tableofcontents

\chapter{Introduction}

\section{Background and Philosophy}
Virtually anyone, who ventured deep enough into the realm of mathematics or computer science, has in one way or another encountered the intriguing duality between simplicity and difficulty. There is perhaps no better example of such duality than the infamous Boolean Satisfiability problem - often abbreviated as SAT or SATISFIABILITY. As will be seen in the coming sections, the SAT problem statement is intriguingly simple, and yet it is arguably the most difficult and most important problem in theoretical computer  science and mathematics being directly associated with one of the Seven Millennium Prize Problems posed by the Clay Mathematics Institute, each of which having a grand prize of one million dollars for whoever manages to solve it. That problem is none other than the \textbf{P vs. NP} question. Such high renown is undoubtedly well-earned given that a solution to that problem has resounding implications on mathematics and many branches of science. 

Informally, the SAT problem asks whether a certain Boolean formula is satisfiable or not; in other words, whether there is a possible truth assignment that makes it true. However, despite the glaring simplicity of the question, it is very hard to determine whether such truth assignment exists in general because the search space grows exponentially with the number of variables in said formula. Accordingly, any exhaustive search algorithm would require exponential run-time in order to be able to solve a general instance of the problem. \newline
\textbf{But how is SAT different from any other difficult problem?} One answer to what makes SAT truly interesting is its completeness relationship over a bigger class of problems called NP problems. Simply speaking, for a problem $A$ to be NP-complete, it means that any problem $B$ belonging to the NP class (i.e. an NP problem) is efficiently reducible or convertible to $A$, which means that a solution to $A$ can be transformed into a solution to $B$. Problems that are at least as hard as NP-complete problems are informally called \textit{intractable}. The formal and rigorous nature of the NP class of problems as well as its relationship to the P problem class shall be later explained in more rigor. However, to give a rough idea, it suffices to say that an NP problem is one whose solutions can be verified easily. For example, a solution to CLIQUE\footnote{CLIQUE stands for the NP-complete Graph Clique Decision Problem where we want to know if a graph G contains a clique larger than an integer k.} can be easily checked by verifying whether the claimed set of vertices really form a clique (and of course, are more than the given integer). Unfortunately, solving the problem might not be nearly as easy as verifying its solutions, and that is the essence of the P vs. NP question. On the other hand, a P problem is one that can be solved easily, hence also verified easily (by just solving it and comparing the results). 

Philosophically, solution and verification are often viewed as analogous to discovery and understanding, respectively. The rationale behind the analogy between discovery and solvability is that most mathematical problems can be framed as computational ones \cite{cook1984can}, and indeed, human thinking can be viewed as an intricate computation. The reason why we do not entirely delegate mathematical discoveries to computers is that as far as we know, discovery is intractable! A similar analogy is often made between understanding and verifiability since understanding a given problem entails being able to verify its proposed solutions. For example, most people understand the problem of solving a Rubik's cube; it is simply arranging the cube such that every side has only one color. It is very easy to distinguish a correctly solved Rubik's cube from otherwise by simply looking at it. However, it is not nearly as easy to solve a scrambled one.

\newpage
\section{An Overview of Turing Machines}
On the crossroad that joins computer science with the realm of mathematics, three main areas reside, namely, \emph{computability}, \emph{complexity}, and \emph{automata}, each of which has certain types of questions with which it concerns itself. Computability theory asks whether a given task is computable (i.e. can be performed by a computer), and complexity theory asks how easy or hard such task is for a computer while automata theory focuses on developing mathematical models of computation. This section aims to serve as a gentle introduction to the most relevant objects and notions of the three areas based on the work of \cite{sipser2012introduction}. 

\subsection{Ordinary Turing Machines}
\label{section:ord_tm}
\begin{figure}[h]
    \centering
    \includegraphics[width=.6\textwidth]{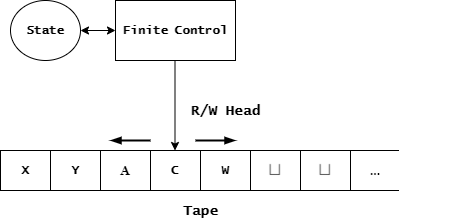}
    \caption{Turing Machine}
    \label{fig:tm}
\end{figure}

As of date, Turing machines are the strongest model for general-purpose computing since their conception by Alan Turing in 1936. Generally, a Turing machine can be described using the following elements:

\begin{itemize}
    \item Infinite memory tape
    \item Machine state
    \item Finite control over state transitions
    \item Read-write tape heads that can read/write from/to the tape
\end{itemize}
These elements can visualized as shown in Figure \ref{fig:tm}.

The tape, which represents the machine memory, can be read or written to at any location by a read-write head that can only move to the left or to the right (by one tape cell) across the tape based on the decision of the finite control. Such decision is determined through the current machine state along with the tape symbol read at the current location. Some states are transitory or intermediate while others are \textbf{halting} states being either \emph{accepting} states or \emph{rejecting} ones. There are only two possible outcomes to the computation performed by a halting Turing machine: either \emph{accept} or \emph{reject} depending on the terminal state it halts on. If no such terminal state is reached, the machine (i.e., program) continues ad infinitum; that is, it does not halt.
\newline

When the program starts, the tape only contains the input symbols on its prefix locations while the rest contains the empty symbol $\sqcup$. For any intermediate information the machine needs to store, it can write it somewhere on the tape and read it again later. Note that the head movement is not necessarily unidirectional throughout the program but can rather apply multiple passes on the same tape locations.

To demonstrate how useful such model can be in describing computation, we shall apply it to language membership problem, that is, verifying whether some input string follows a certain language pattern. More formally, let language $L = \{s\#s \mid s \in \{0,1\}^*\}$ and string $w$ be an input string, and we wish to decide whether $w \in L$. That is, we want to know whether string $w$ is composed of two identical binary strings separated by a \# character. A straightforward approach to do such processing would be to compare each character with the one at the congruent location across the \# character and check if they match or not and that one \# exists. Each processed character aside from \# gets crossed off by the x symbol. Thus, we can describe a Turing machine $M$ to perform this task as follows:
\newline

\begin{example}
\label{example:TM1}
$M:= \text{ On input string }w:$ 
\begin{enumerate}
    \item For each character $c_i \in w$ on the left side of \#, cross it and slide tape head to the right until \# is encountered. If a \# symbol is found, slide to the first uncrossed $c_j$ symbol after the last x location. If $c_i=c_j \Rightarrow$ cross off $c_j$ with the 'x' symbol, else $\Rightarrow$ \emph{reject}. If no \# is found, also \emph{reject}.
    \item Check if there are any remaining symbols aside from \# and x. If there are, \emph{reject}, otherwise, \emph{accept}.
\end{enumerate}

Figure \ref{fig:lang_mem_tm} is one showcase of $M$ on a sample input $w=101\#101$. Note that it does not show all intermediate steps.
\begin{figure}[h]
    \centering
    \includegraphics[width=.5\textwidth]{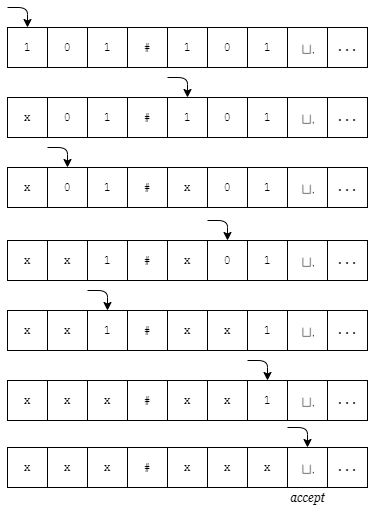}
    \caption{Language Membership Turing Machine}
    \label{fig:lang_mem_tm}
\end{figure}
\end{example}

To formalize the control transition from one state to another, we define a transition function $\delta$ that given the current state and currently read symbol on the tape, determines the next move. To achieve this, we define the following sets.

\begin{tcolorbox}[title=Turing Machine Set Denotations]
\label{tm_sets}
For a Turing machine, we denote three finite sets $Q, \Sigma, \Gamma$ as:
\begin{enumerate}
    \item $Q$ is the set of possible TM states.
    \item $\Sigma$ is the input alphabet excluding the blank symbol $\sqcup$.
    \item $\Gamma$ is the tape alphabet, where $\sqcup \in \Gamma$ and $\Sigma \subseteq \Gamma$.
\end{enumerate}
\end{tcolorbox}

As such, we can define the $\delta$ function in the form: $Q\times\Gamma \rightarrow Q\times\Gamma\times\{L,R\}$. To elaborate, $\delta(q,a)=(r,b,L)$ means that when the machine is at state $q$ and the head reads symbol $a$ from the tape, the next move would be to transition to state $r$, write symbol $b$ on the tape (replacing $a$), and move the head one location to the left. Now that we have all the formal objects described, we can formally define a Turing machine.

\begin{definition}
A \textbf{Turing machine} is a 7-tuple ($Q, \Sigma, \Gamma, \delta, q_0, q_{accept}, q_{reject}$) where $Q, \Sigma, \Gamma$ are finite sets such that:
\begin{enumerate}
    \item $\delta: Q\times\Gamma \rightarrow Q\times\Gamma\times\{L,R\}$
    \item $q_0 \in Q$ is the initial state.
    \item $q_{accept} \in Q$ is the accepting state.
    \item $q_{reject} \in Q$ is the rejecting state.
\end{enumerate}
\end{definition}

Based on that definition, we can formally describe the computation for Example \ref{example:TM1}. First, we define our machine sets:

\begin{center}
$\Sigma = \{0,1\}$ \\
$\Gamma = \{0, 1, \text{x}, \sqcup\}$ \\
\end{center}

\begin{center}
$Q = \{A, B, D, \text{accept}, \text{reject}\}\bigcup \;\{F_s\;|\:s \in \Gamma\}\bigcup \;\{C_s\;|\:s \in \Gamma\} \text{ where:}$
\end{center}

\begin{table}[h!]
\centering
 \begin{tabular}{||c | c||} 
 \hline
 $Q$ & \textbf{Meaning}  \\ [0.5ex] 
 \hline\hline
 $A$ & $\text{capture first uncrossed input symbol to the left of the \#}$  \\ \hline
 $F_{s}$ & $\text{search forwards for the \# symbol after capturing symbol }s$  \\ \hline
 $B$ & $\text{search backwards for the \# symbol}$  \\ \hline
 $C_s$ & $\text{match first uncrossed input symbol to the right of the \# with symbol }s$  \\ \hline
 $D$ & $\text{find first x symbol to the left of the \#}$  \\ \hline
 $\emph{accept}$ & $\text{accepting state}$  \\ \hline
 $\emph{reject}$ & $\text{rejecting state}$  \\ [1ex] 
 \hline
 \end{tabular}
 \caption{Example \ref{example:TM1} TM States Interpretations}
\end{table}

We then specify the TM's initial state as $q_0 = A$. Now let's partially define the delta function mappings for some of the machine configurations of interest.
Note that Table \ref{table:delta_func} only includes the important configurations that test the correctness of the TM in light of Example \ref{example:TM1}.

\begin{table}[h!]
\centering
 \begin{tabular}{||c | c||} 
 \hline
 $Q\times\Gamma$ & $\delta(Q\times\Gamma)$ \\ [0.5ex] 
 \hline\hline
 ($A$, 1) & ($F_{1}$, x, R) \\ \hline
 ($F_{1}$, 0) & ($F_{1}$, 0, R) \\ \hline
 ($F_{1}$, 1) & ($F_{1}$, 1, R) \\ \hline
 ($F_{1}$, \#) & ($C_1$, \#, R) \\ \hline
 ($F_{1}$, $\sqcup$) & ($\emph{reject}$, $\sqcup$, R) \\ \hline
 ($C_1$, 1) & ($B$, x, L) \\ \hline
 ($C_1$, 0) & ($\emph{reject}$, 0, L) \\ \hline
 ($B$, x) & ($B$, x, L) \\ \hline
 ($B$, \#) & ($D$, \#, L) \\ \hline
 ($D$, 1) & ($D$, 1, L) \\ \hline
 ($D$, x) & ($A$, x, R) \\ \hline
 ($A$, \#) & ($C_{\#}$, \#, R) \\ \hline
 ($C_{\#}$, $\sqcup$) & ($\emph{accept}$, $\sqcup$, L) \\ \hline
 ($C_{\#}$, $\#$) & ($\emph{reject}$, $\#$, L) \\ \hline
 ($C_{\#}$, 0) & ($\emph{reject}$, 0, L) \\ \hline
 ($C_{\#}$, 1) & ($\emph{reject}$, 1, L) \\ [1ex] \hline
 \end{tabular}
 \caption{Example \ref{example:TM1} Partial $\delta$ Mapping}
 \label{table:delta_func}
\end{table}

The notion of Turing machines is also helpful because, besides providing a general model for computation, it gives a sense of the time required to perform the computation by comparing how the number of control moves (i.e. steps) grows as input size increases, which is typically referred to as the time complexity of such TM.

What we described in Example \ref{example:TM1} is called deciding a language, that is, deciding whether a word belongs to some language. Framing computational problems as language membership problems is, in fact, much more powerful than it seems, as we will see later. Now let's define what it means to decide a language:

\begin{definition}
A language $L$ is Turing-decidable if there exists a Turing machine $M$ that \emph{accepts} on all $\{w\mid w \in L\}$ and \emph{rejects} on all $\{w\;|\:w \notin L\}$. 
\end{definition}
\begin{remark}
In order for $M$ to decide $L$, $M$ needs to halt on all inputs.
\end{remark}

On that basis, any language that cannot be decided with an ordinary Turing machine is called \emph{Turing-undecidable} or \emph{undecidable}. One of the most famous undecidable problems, and the first to be proven so, is the notorious \textbf{Halting Problem} formally denoted as $\text{HALT}$:

\begin{center}
    $\text{HALT}=\{\langle M, w \rangle \;|\: M \text{ is a TM that halts on input }w\}$
\end{center}

Alan Turing, in 1936, proved by contradiction that $\text{HALT}$ is undecidable. This undecidability of HALT was one of the most important theorems for theoretical computer science and mathematics because it definitively proved that some parts of mathematics cannot be decided. That is, some mathematical questions may never be answered by a consistent system of thought. For the sheer simplicity of Turing's proof for such a seemingly difficult and crucial question, we shall present it here.

\begin{theorem}
HALT is Turing-undecidable.
\end{theorem}
\begin{proof}
For the sake of contradiction, assume there is a Turing machine \textbf{H} that can decide the halting problem. That is, given a TM's description \textbf{A} along with its input \textbf{w}, machine H determines whether or not A halts on input w as shown in Figure \ref{fig:halt}.

\begin{figure}[h]
    \centering
    \includegraphics[width=.3\textwidth]{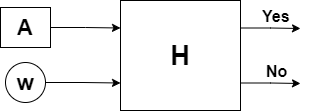}
    \caption{A TM That Decides HALT}
    \label{fig:halt}
\end{figure}

Now we construct a new Turing machine $\text{H}^*$ by slightly modifying H to make it go into an infinite loop when input machine A halts on input w (i.e., when H gives a Yes outcome) as illustrated in Figure \ref{fig:halt2}.

\begin{figure}[h]
    \centering
    \includegraphics[width=.4\textwidth]{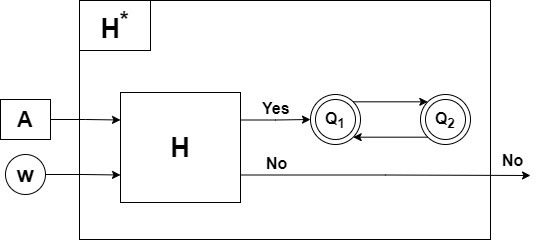}
    \caption{Modified Encapsulating Machine $\text{H}^*$}
    \label{fig:halt2}
\end{figure}

As such, upon receiving a halting input, $\text{H}^*$ can either reject (i.e., output No) or loop forever (i.e., not halt) since it goes into an infinite back-and-forth transition between two states $\text{Q}_1$ and $\text{Q}_2$. Now what if we insert the $\text{H}^*$ machine description as both input machine A and input string w to $\text{H}^*$? Here, we are interested in the output of the internal H machine for which there can be only two cases: either Yes or No. 

\begin{enumerate}[label=\textbf{Case \arabic*:}, leftmargin=4em]
    \item H outputs \textbf{Yes}: that immediately implies that $\text{H}^*$ halts on $\text{H}^*$, but that would mean that $\text{H}^*$ loops forever since it definitionally does that when its internal H outputs Yes. This is a contradiction since it means that $\text{H}^*$ both halts and does not halt.
    \item H outputs \textbf{No}: that immediately implies that $\text{H}^*$ does not halt on $\text{H}^*$, but that would mean that $\text{H}^*$ halts with a No since it propagates the output of its internal H when it outputs No. This is a contradiction since it means that $\text{H}^*$ both halts and does not halt.
\end{enumerate}

Therefore, there cannot exist a Turing machine that decides HALT since its very existence is contradictory. Hence, HALT is undecidable.
\end{proof}

In adjacency to Turing-decidability, there lies another important and more general concept of \emph{Turing-recognizability}:
\begin{definition}
A Turing machine $M$ recognizes a language $L$ if  $M$ \emph{accepts} on all $\{w\;|\:w \in L\}$, but does not necessarily halt or \emph{reject} on all $\{w\;|\:w \notin L\}$. 
\end{definition}
\begin{remark}
If $M$ decides $L$, then $M$ recognizes $L$.
\end{remark}

\subsection{Multitape Turing Machines}
Typically, when we say \emph{Turing machine}, it is meant as something somewhat more specific that is \emph{deterministic single-tape Turing machine}. Turing machine determinism will be deferred to the next subsection. As the name suggests, a \emph{multitape Turing machine} can have more one than one tape each with their own read-write heads. As such, to have a more general definition that covers a variation of $k$ tapes, we need to modify the control function $\delta$ as:
\begin{center}
$\delta: Q\times\Gamma^k \rightarrow Q\times\Gamma^k\times\{L,R\}^k$
\end{center}
In this definition, a $k$-tuple is used to represent the symbols read on all $k$ tapes and another $k$-tuple to represent the symbols to be written to each tape in the next step. The last $k$-tuple represents the movement of each of the $k$ tapes.
\begin{theorem}
Every multitape Turing machine has an equivalent single-tape Turing machine.
\label{them:multitape-eq}
\end{theorem}
\begin{proof}
We present a way to simulate multiple tapes using a single tape. Let $M$ be a multi-tape TM with $k$ tapes, and $S$ be a single-tape TM. We can then simulate $M$ using $S$ by constructing its tape as the concatenation of all $k$ tapes' contents delimited by the \# symbol. The heads locations would be simulated by dotting the symbols to which they point, that is, replacing each symbol $a$ pointed to by some tape head with symbol $\dot{a}$. Note that \# $\notin \Sigma$ and $\{\dot{a}\;|\: a\in\Sigma\}\cap\Sigma=\phi$. In other words, the special characters are not part of the input alphabet to avoid confusion with input symbols.
\end{proof}

\begin{figure}[h]
    \centering
    \includegraphics[width=.6\textwidth]{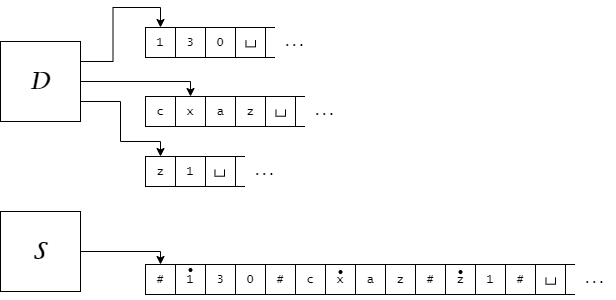}
    \caption{Multitape to Single-Tape Conversion}
    \label{fig:multitape_conv}
\end{figure}


\subsection{Nondeterministic Turing Machines}
Despite the fact that the term \emph{nondeterminism} typically conveys uncertainty, in the context of computation, nondeterminism is a stronger notion than deterministic computation. In essence, a deterministic machine is a nondeterministic one but with only one possible state transition from any given state and input. Both models are similar in their construction with the only difference being the transition $\delta$ function. For a nondeterministic Turing machine (NTM), the transition function in defined as:
\begin{center}
$\delta: Q\times\Gamma^k \rightarrow \mathcal{P}( Q\times\Gamma^k\times\{L,R\}^k)$
\end{center}
where $\mathcal{P}(S)$ is the power set of $S$, that is, the set of all subsets of $S$. As such, a nondeterministic Turing machine branches in a tree-like manner, simultaneously considering more than one possibility at a given state. NTM computation accepts when any of its branches accepts and rejects when all of its branches reject; otherwise, computation never halts. What is clear from such description of NTMs is that an NTM is more computationally powerful or efficient than a deterministic TM. What might not be very clear, however, is that there is an equivalence between the two.

\begin{theorem}
Every nondeterministic Turing machine has an equivalent deterministic Turing machine.
\end{theorem}

\begin{tcolorbox}[title=Proof Idea]
The idea is to simulate a nondeterministic Turing machine $N$ by searching its computation tree using a deterministic Turing machine $D$, and if $D$ finds an accepting state, it accepts, and if it exhausts the entire search tree without finding such state, it rejects. Otherwise, $D$ never halts. In $N$'s computation tree, every node corresponds to some machine configuration\footnote{Machine configuration represents machine state, tape contents, and head position} which can either be accepting, rejecting, or intermediate. Conducting this search in a breadth-first manner is more prudent than depth-first as some branch might not halt while some other branch reaches an \textit{accepting} state, in which case, $D$ might fail to find the accepting node as it gets stuck trying to cover the depth of the infinite branch.
\end{tcolorbox}

\begin{proof}
We construct $D$ as a 3-tape TM which, based on theorem \ref{them:multitape-eq}, has an equivalent single-tape TM. The three tapes of $D$ shall be used as follows:
\begin{itemize}
    \item \textbf{Tape 1 (input)}: stores the input tape and is never modified.
    \item \textbf{Tape 2 (simulation)}: keeps a copy of $N$'s tape on some configuration node.
    \item \textbf{Tape 3 (address)}: tracks $D$'s location on $N$'s computation tree which is needed for the BFS traversal.
\end{itemize}

\begin{figure}[h]
    \centering
    \includegraphics[width=.6\textwidth]{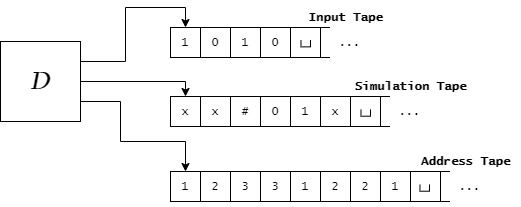}
    \caption{Deterministic TM $D$}
    \label{fig:ntm-to-tm}
\end{figure}

One pithy way to represent the track location on tape 3 would be as a sequence representing the child order relative to its parent at each level starting from root. For instance, node 421 represents the node we reach by going from root onto its $4^{th} child \rightarrow 2^{nd} child \rightarrow 1^{st} child$. However, as with any tape, it needs to have a finite alphabet, so we should have a limited set of possible integers in that sequence. In other words, a child's order should be bounded; therefore, we define tape 3's alphabet as $\Sigma=\{1,2,...,b\}$ where $b$ is the maximum branching factor for $N$'s computation tree. It is worth noting that the address on tape 3 could be invalid if it corresponds to a nonexistent child. With that settled, we can now start describing $D$ as follows:

\begin{enumerate}[label=\textbf{Step \arabic*:}, leftmargin=4em]
    \item Tape 1 contains input $w$ while tapes 2 \& 3 are empty.
    \item Copy tape 1 to tape 2.
    \item Simulate the current computation node using tape 2 after consulting the next symbol on tape 3 at each step to make the next allowable transition. If the transition pointed to by the symbol is invalid (i.e., does not correspond to a valid $\delta$ transition) or if no symbols are left on the tape, then abort that simulation step and go to step 4. If an accepting configuration is found, \emph{accept} the input. If a rejecting configuration is instead found, go to step 4.
    \item Modify the string on tape 3 to have the lexographically next string, then simulate the corresponding branch by going to step 2.
\end{enumerate}

\noindent Note that $D$ repeats steps it already performed before because it traverses (i.e., simulates) the computational tree from the root till the last child on the node address for every new node address. In essence, $D$ performs BFS by a DFS dynamic, that is, it dives through the depth of the tree to reach a certain node and then returns to the root (step 2) before going for the next node in the BFS order. Note that this is not a precise description of the simulator Turing machine since rigorously describing it would be cumbersome and does not serve the core purpose of this proof. However, moderate understanding of how Turing machines work suffices to see that it is possible to construct such machine as $D$.

\end{proof}

In spite of this functional equivalence between NTMs and DTMs, it is generally believed that they differ in their time complexities with NTMs being on the much lower end, taking polynomial upper bounds to solve problems that the best DTMs can only solve in exponential time. However, before any optimistic ideas arise, NTMs cannot be implemented by either classical computers or even quantum computers. Instead, NTMs are usually employed to classify certain problems based on their computability by an NTM (more on that in coming sections). It is also worth mentioning that NTMs abide by Turing-decidability just like DTMs do, so an NTM cannot solve the halting problem, for instance, since it is undecidable.

\section{Preliminaries}

\begin{definition}
A Boolean clause is a conjunction or disjunction of literals where a literal is a Boolean variable or its negation. The default usage of \say{clause} is to mean disjunctive clause.
\end{definition}

\begin{definition}
An alphabet $\Sigma$ is a set of symbols used to encode information such as inputs and proofs.
\begin{remark}
The most common encoding is binary encoding on the alphabet $\{0,1\}$.
\end{remark}
\end{definition}

\begin{definition}
A language is a set $L \subseteq \Sigma^*$ where $\Sigma^*$ is the set of all finite strings on the alphabet $\Sigma$. 
\end{definition}

\begin{definition}
A decision problem is some language $L \subseteq \Sigma^*$ which represents the set of all inputs that satisfy the underlying predicate of the problem.

\begin{remark}
Another way to look at these inputs is that they are the only ones on which a Turing machine that solves $L$ \textit{accepts}.
\end{remark}
\end{definition}

\begin{convention}
The outcome (i.e., decision) of a Turing machine $M$ on some input $x$ is denoted as $M\langle x \rangle$. We write $M\langle x \rangle = 1$, Yes, or \textit{accept} to mean that $M$ \textit{accepts} on input $x$. Conversely, we write $M\langle x \rangle = 0$, No, or \textit{reject} to mean that $M$ \textit{rejects} on input $x$.
\end{convention}

\begin{convention}
When a Turing machine $M$ is expected to have an output string aside from its halting decision on input $x$, we denote that output as $M(x)$. Structurally, that output can be considered as the tape contents at the halting state.
\end{convention}

\begin{convention}
The terms \emph{decision problem} and \emph{language} are often used interchangeably with \emph{language} being more frequent in abstract contexts.
\end{convention}

\begin{definition}
A Turing machine $M$ is PTIME if there exists a polynomial $p$ such that $M$ takes $p(|x|)$ steps (i.e. tape head moves) to compute $M\langle x \rangle$ where $x\in\Sigma^*$.
\end{definition}

\begin{definition}[Big-O Notation]
For two functions $f$ and $g$, we write $f(n) = \mathcal{O}(g(n))$ \emph{iff} there exists:
\begin{enumerate}
    \item $n_0 \in \mathbb{Z}^+$
    \item $c \in \mathbb{R}^+$
\end{enumerate}
such that $\forall n \ge n_0 (f(n) \le cg(n))$. This expresses an upper bound.
\end{definition}

\begin{definition}[Big-Omega Notation]
For two functions $f$ and $g$, we write $f(n) = \Omega(g(n))$ \emph{iff} there exists:
\begin{enumerate}
    \item $n_0 \in \mathbb{Z}^+$
    \item $c \in \mathbb{R}^+$
\end{enumerate}
such that $\forall n \ge n_0 (f(n) \ge cg(n))$. This expresses a lower bound.
\end{definition}

\begin{definition}[Big-Theta Notation]
Function $f(n) = \Theta(g(n))$ \emph{iff} $f(n) = \Omega(g(n))$ and $f(n) = \mathcal{O}(g(n))$. That is, $g(n)$ is a tight bound for $f(n)$.
\end{definition}

\begin{convention}
For a problem $L$, a bound function $\mathcal{B} \in \{\mathcal{O}, \Omega, \Theta\}$, and a function $f$, we write $L=\mathcal{B}(f(n))$ if there exists a TM $A$ with run-time function $T$ such that $T(n)=\mathcal{B}(f(n))$.
\end{convention}

\chapter{Problem Hardness}
\begin{fquote}[Alan Turing, 1939][The Father of Computer Science]
We shall not go any further into the nature of this oracle apart from saying that it cannot be a machine.
\end{fquote}

Prior to the conception of theoretical complexity, mathematicians had no rigorous basis to determine the difficulty level of a given problem; instead, they merely followed a rough sense of how hard they struggled with it. Fortunately, the relatively recent branch of theoretical complexity takes it upon itself to try to set formal grounds and boundaries to computational problem hardness relative to other problems. It bases this hardness comparison between two problems on the number of steps\footnote{We usually do not care about the exact number of steps but rather the type of the run-time function itself (e.g., polynomial, exponential, etc).} it takes a Turing machine to transform one of them to the other. For simplicity, we can treat a TM as a program or an algorithm designed to solve a specific problem without paying much attention to its internals. This leads us to the concept of complexity classes.

\section{Complexity Hierarchy}
\label{section:comp_hier}
It is worth noting that \emph{complexity} in our scope refers to the amount of computational resources (i.e. time and/or space) needed to solve a problem. The relationship between problem input size and the resources needed to solve it essentially determines the complexity class to which the problem belongs. Famous complexity classes include PTIME (aka \textbf{P}), PSPACE, NPTIME (aka \textbf{NP}), and EXPSPACE.

In this paper, we will be focusing on only the two most prominent: P and NP. The P class refers to problems that are deterministically \emph{solvable} in polynomial time; that is, the time needed to decide them by a TM is a polynomial function in the input size. This is also known as being polynomially solvable. On the other hand, the NP class refers to problems that are polynomially \emph{verifiable}. That  means that their decisions have proofs that are polynomial in length (w.r.t. input size), and the time needed to verify their solutions (i.e., proofs of decision correctness) by a TM is a polynomial function in the solution or proof size. 

\begin{definition}
A decision problem $L \subseteq \Sigma^*$ is in P if there is a PTIME algorithm (i.e., Turing machine) M such that $x \in L$ $\iff$ $M\langle x \rangle  = 1$.

\begin{remark}
In that sense, $M$ is a solver algorithm as it determines the outcome of the decision problem (Yes or No).
\end{remark}
\end{definition}

\begin{definition}
A decision problem $L \subseteq \Sigma^*$ is in NP if there exist a polynomial $p$ and a PTIME algorithm (i.e., Turing machine) $V$ such that $x \in L$ $\iff$ $\exists~ y \in \Sigma^{\le p(|x|)}(V\langle x,y \rangle = 1)$.

\begin{remark}
Here, $V$ is a verifying algorithm as it takes $y$ as a proof string that $x \in L$ and checks if it is a valid proof. $\Sigma^{\le p(|x|)} = \{w: |w| \le p(|x|)\}$ where $|w|$ is the length of string $w$, which ensures the proof length is polynomially bounded in size.
\end{remark}
\end{definition}

To give an example of decision proofs, in a general SAT problem, a proof string $y$ would be an interpretation (i.e., variable assignment) that is claimed to satisfy the given formula.  \newline
Let $\phi = (x_1 \land x_2 \land \neg x_3)$, then $y = 110$ is a valid proof string that $\phi \in$ SAT since that assignment satisfies $\phi$. A proof strings is also sometimes called a \emph{certificate} or \textit{witness}. This concept of polynomially verifiable proof strings is often invoked when analyzing NP problems in the deterministic sense (i.e., using DTMs). One common misconception about the NP class is that NP stands for "\textbf{N}on-\textbf{P}olynomial" while, in fact, it stands for \say{\textbf{N}on-deterministic \textbf{P}olynomial.} That naming is based on another definition that is equivalent to the previous one, but relies on NTMs instead.

\begin{theorem}
A decision problem $L \subseteq \Sigma^*$ is in NP if and only if it can be solved by a nondeterministic Turing machine (NTM) in polynomial time.
\end{theorem}
\begin{proof}
$(\Rightarrow):$ if $L \in \text{NP}$, then the size of its proof certificate is polynomially bounded, so all of its possibilities can be exhausted by an NTM in polynomial time through exhaustive branching. And since each branch, corresponding to one possibility of a certificate, can be deterministically verified (i.e., through a DTM) in PTIME, then the entire certificate search process can be done by an NTM in PTIME. If a valid proof certificate is found, then the NTM \textit{accepts}, otherwise, it \textit{rejects}.

$(\Leftarrow):$ if $L$ can be solved by an NTM $N_L$ in PTIME, then certificates can be set up as the set of nondeterministic choices that $N_L$ makes until it reaches an accepting state (if any). Note that if $N_L$ \textit{accepts}, there must be at least one such polynomially bounded certificate since certificate length would be determined by the depth of the accepting branch in $N_L$. We can then devise a deterministic TM $V_L$ that can verify such certificates by simulating the $M_L$ computation tree on the choices described by the certificate and checking whether they result in an accepting node (i.e., configuration). For instance, the witness $421$ represents the node we reach by going from root onto its $4^{th} child \rightarrow 2^{nd} child \rightarrow 1^{st} child$, which may or may not be an accepting node.
\end{proof}

In the literature, the words "efficiently" (\say{easily}, or \say{feasibly}) and \say{polynomially} are sometimes used interchangeably, and likewise with \say{infeasibly} and \say{superpolynomially} (e.g., exponentially). That convention is in accordance with the Feasibility Thesis \cite{cook1990computational}:

\begin{namedthm}{Feasibility Thesis}
A natural problem has a feasible algorithm iff it has a polynomial-time algorithm.
\end{namedthm}
We call a problem \textit{natural} when it arises from a genuine objective we need to accomplish in real-life practice as opposed to a problem that was conceptualized merely for the purpose of proving a point. An example of the former would be sorting sequences while an example of the latter would be producing an arbitrary string of length $n^{1000}$. The latter problem indeed has a polynomial-time algorithm despite being totally infeasible, but we do not care about it since it does not arise in a practically useful context.
Though the statement of the Feasibility Thesis is largely subjective, it applies well on almost all problems known to the literature.
As such, we can generally define P problems as feasibly solvable and NP problems as feasibly verifiable. Accordingly, as mentioned before, $\text{P} \subseteq \text{NP}$ since it is easy to verify that which is easily solvable. 
\begin{lemma}
$\text{P} \subseteq \text{NP}$
\end{lemma}

\begin{figure}[h]
    \centering
    \includegraphics[width=.3\textwidth]{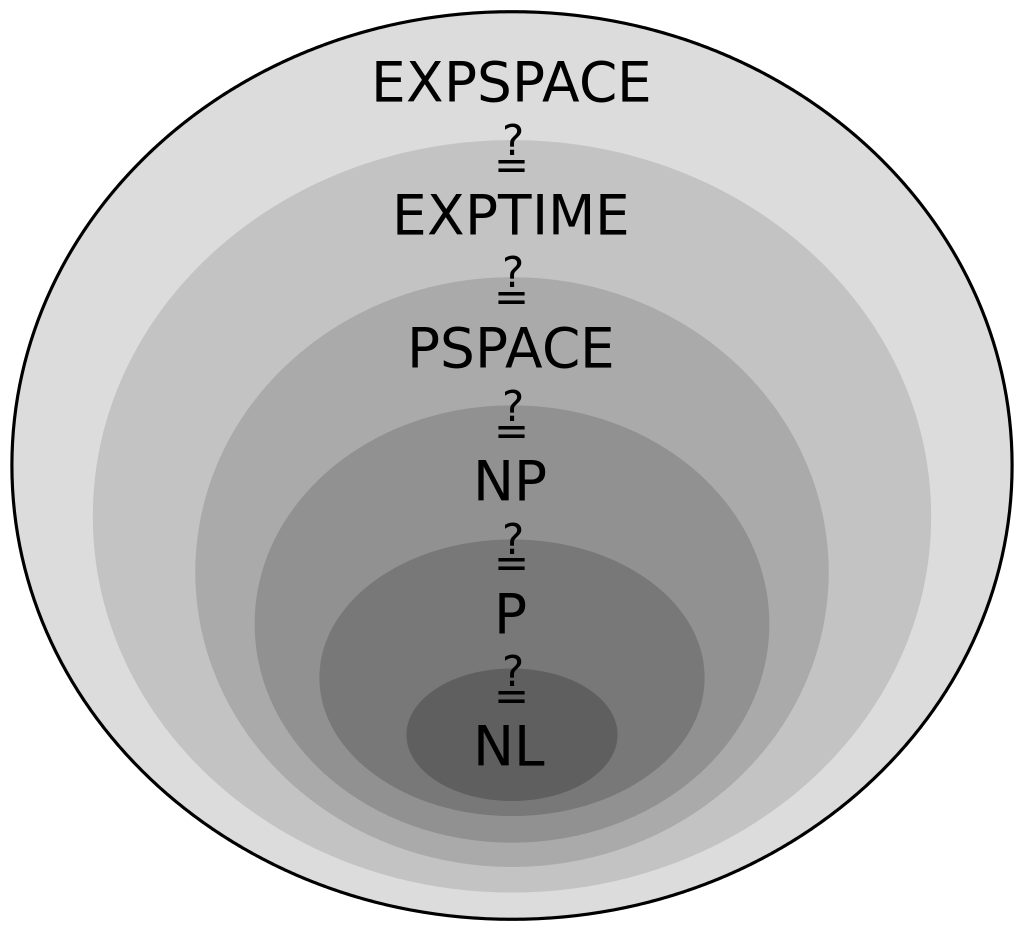}
    \caption{General Complexity Class Hierarchy}
    \label{fig:comp_hier}
\end{figure}

\begin{proof}
Let a problem $L \in P$. Then, there exists a PTIME algorithm M such that $x \in L$ \emph{iff} $M(x) = 1$. We can then construct a PTIME verifier machine $V$ as: $V\langle x,y \rangle = M\langle x \rangle$. In other words, we can ignore the proof $y$ and just solve $L$ using $M$ in PTIME directly. Thus, $L \in NP$
\end{proof}

Our discourse in this section may suggest that problem classes can generally be split into two main umbrellas: \emph{deterministic} (e.g., PTIME, LOGSPACE) and \emph{non-deterministic} (e.g., NPTIME, NLOGSPACE) based on the type of Turing machines associated with them. However, such categorization says little about certain classes being harder than others, which is why problem complexity hierarchy are typically viewed as subsets of others like in Figure \ref{fig:comp_hier}. The question marks on the figure above allude to the fact that it is not yet known whether some of them are equal to each other or not, as a certain problem is classified under one class when a solver algorithm exists - or known to exist - for it which has a time complexity belonging to that class. Of course, oblivion of such algorithm does not preclude its existence which is why the fact that no polynomial-time algorithm was found to solve a certain problem like SAT does not mean that SAT is not in P; we just do not know that yet.

\newpage
\section{Problem Reduction}
In this section, we introduce a very important notion to analyze problem complexity (or hardness) and that is through the ability to transform or reduce a given problem to another. Following the same convention used with solvability and verifiability, when we say \say{reduce}, we mean \say{reduce in polynomial time} or \emph{polynomially reduce}. \newline

To reduce problem $A$ to problem $B$ is to convert input of $A$ to an input of $B$ such that an answer to the $B$ problem instance automatically gives us an answer to the $A$ instance. The fact that $A$ is reducible to $B$ is formally expressed as $A \leq_{\mathcal{P}} B$.

\begin{definition}
For two decision problems $A,B$, we write $A \leq_{\mathcal{P}} B$ if there exists a PTIME TM $\mathcal{F}$ such that for any input $x$ to problem A, $\mathcal{F}(x) \in B$ $\iff$ $x \in A$. 
\end{definition}

To say that $A \leq_{\mathcal{P}} B$ is to say that $A$ is at most as hard as $B$, which is intuitive, for $A$ cannot be harder than $B$ as long as answering $B$ promptly grants us an answer to $A$. In that context, $\mathcal{F}$ is a transformer algorithm or a machine that translates an instance of $A$ to an instance of $B$.
While this definition of reduction is for decision problems, the general idea for any two problems is similar, but this paper mainly focuses on decision problems. One simple example of a problem reduction is reducing the problem of finding the minimum element in a sequence $X$ of integers (problem $A$) to the problem of sorting a sequence $X$ of integers ascendingly (problem $B$). The reduction here is trivial because we can pass on the same sequence $X$ as input to problem $B$, then we take the very first element in the output of $B$, which would be the answer to $A$. Note that none of these two problems are decision problems because their answer is not a binary Yes or No. 

It is now perhaps apparent how useful the notion of problem reduction is to get a bounded sense of how computationally hard (or easy) a problem is by transforming it into another. What makes this of even greater use is that polynomial reducibility is transitive, allowing us to relate entire classes of problems using a single reduction instance between any two of their members.

\begin{lemma}
If $A \leq_{\mathcal{P}} B$ and $B \leq_{\mathcal{P}} C$, then $A \leq_{\mathcal{P}} C$.
\end{lemma}

\begin{proof}
Let $\mathcal{F}$ be an $A$-to-$B$ reducer, and $\mathcal{H}$ be a $B$-to-$C$ reducer. Then, $x \in A \iff \mathcal{F}(x) \in B$, and  $y \in B \iff \mathcal{H}(y) \in C$. Therefore, it follows that $\mathcal{H}(\mathcal{F}(x)) \in C$ \emph{iff} $x \in A$; that is, $\mathcal{H} \circ \mathcal{F}$ is an $A$-to-$C$ reducer. As such, it remains to show that $\mathcal{H} \circ \mathcal{F}$ is PTIME, but since polynomials are closed under composition\footnote{The composition of two polynomials is a polynomial.}, then $\mathcal{H} \circ \mathcal{F}$ is PTIME as well. Hence, $A \leq_{\mathcal{P}} C$.
\end{proof}

In order to give a closer computational analysis of a given problem, complexity bound analysis is used to narrow down the time (or space) complexity of said problem. As such, there are three popular types of asymptotic bounds: lower (best-case), upper (worst-case), and exact. Another widely used complexity analysis is average-case complexity. To relate back the complexity bounds notations, we explore the two complexity bound implications of problem reducibility in the following lemmas:

\begin{corollary}
\label{red_and_comp1}
If $A \leq_{\mathcal{P}} B$ and $A = \Omega(f(n))$, then $B = \Omega(f(n))$.
\end{corollary}

\begin{corollary}
\label{red_and_comp2}
If $A \leq_{\mathcal{P}} B$ and $B = \mathcal{O}(f(n))$, then $A = \mathcal{O}(f(n))$.
\end{corollary}
It would be somewhat superfluous to tackle the proofs of both corollaries as they can be semantically inferred from the conceptual meaning of reducibility, worst-case, and best-case complexities. To explain, for corollary \ref{red_and_comp1}, if $A$ is at most as hard as $B$, and $A$ has a lower bound complexity $f(n)$, then $B$ must have that lower bound as well\footnote{Note that lower and upper bound complexities do not need to be asymptotic.}. That is because otherwise, $B$ would have a lower bound than $\Omega(f(n))$, in which case there must exist at least one input word $w$ such that it is easier to decide if $w \in B$ than deciding if $w_A$\footnote{$w_A$ denotes word $w$ after transforming it to an input word recognizable for language $A$, that is, by a $B$-to-$A$ reducer.} $\in A$, which contradicts that $A$ is not harder than $B$. The same method can be used for corollary \ref{red_and_comp2} by showing that $A \neq \mathcal{O}(f(n))$ would imply that $A$ is sometimes harder than $B$.

What we have presented thus far is, in fact, a special case of a more general notion of reduction, and that is \textbf{\emph{Turing reducibility}} which places no restrictions on the time complexity of the reduction TM. It instead relies on the construct of an \textbf{\emph{oracle}}.

\begin{definition}
An \textbf{\emph{oracle}} is a black box device capable of deciding a language in a single step. That is, for any word $w$, the oracle for language $A$ can instantly decide whether $w \in A$.
\end{definition}

In that sense, an oracle for language $A$ is a magical box that takes in a word $w$ and immediately decides if $w \in A$, and the way it determines that is irrelevant since oracles are purely abstract analytical tools used to establish other more useful concepts. Furthermore, when an oracle is attached to a Turing machine, the resultant is called an \textbf{\emph{oracle Turing machine}}.

\begin{definition}
An \textbf{\emph{oracle Turing machine}} $M^A$ is a modified Turing machine that can consult an oracle for language $A$.
\end{definition}

It is well worth noting that an oracle can even solve undecidable problems like the \emph{Halting Problem} for ordinary Turing machines. Remember that for a problem to be \textit{decidable}, there must be an ordinary Turing machine that can decide it, which is different from being decided by an oracle or an oracle TM. We are now ready to define Turing reducibility:

\begin{definition}
For two languages $A, B$, we write $A \leq_T B$ to denote that $A$ is decidable relative to $B$. That is, there exists an oracle Turing machine $M^B$ that can decide $A$.
\end{definition}

\begin{theorem}
If $A \leq_T B$ and $B$ is decidable, then $A$ is decidable.
\end{theorem}
\begin{proof}
Let $M^B$ be an oracle TM that decides $A$, and $D$ be an ordinary TM that decides B. Then, we can replace the $B$-oracle in $M^B$ by $D$, which will result in a compound \emph{ordinary} TM that decides $A$. Thus, $A$ is decidable.
\end{proof}

This notion of reducing or deciding a problem \emph{relative} to another is essentially the central concept of problem reduction that is often denoted as \emph{\textbf{relativization}}.

\begin{lemma}
If $A \leq_{\mathcal{P}} B$, then $A \leq_T B$.
\end{lemma}
\begin{proof}
Let $\mathcal{F}$ be an $A$-to-$B$ reducer TM. We can build an oracle Turing machine $\mathcal{F}^B$ that decides $A$ as follows:

\noindent$\mathcal{F}^B:= \text{ On input string }x$:
\begin{enumerate}
    \item Use $\mathcal{F}$ to convert $x$ to $x_B$, an input for language $B$.
    \item Query the $B$-oracle for string $x_B$: if it accepts, \emph{accept} $x$, otherwise, \emph{reject} $x$. 
\end{enumerate}
Therefore, there exists an oracle Turing machine $\mathcal{F}^B$ that decides $A$.
\end{proof}

\newpage
\section{NP-Completeness}
We have introduced the complexity class NP followed by the concept of polynomial-time reductions\footnote{For simplicity, we shall just called them \emph{reductions}.}. In this section, we present NP-completeness as a property of languages (i.e. decision problems) that combines both concepts. 

\begin{definition}
A language $L$ is NP-complete if: 
\begin{enumerate}
    \item $L \in$ NP.
    \item Every language in NP must be polynomially reducible to $L$.
\end{enumerate}
\end{definition}

In that sense, NP-complete problems are the hardest problems in NP. However, that does not mean they are the hardest in general since there are problems much harder than NP-complete problems, which directly implies these harder problems cannot be in NP. Hence, these harder problems do not necessarily satisfy the first condition for NP-completeness, and these are called NP-hard problems which means problems at least as hard as NP-complete problems. To give perspective, recall we defined NP problems as ones that are polynomially verifiable on a certificate. One example of such problem is CLIQUE:
\begin{center}
    $\text{CLIQUE} = \{\langle G,k \rangle \;|\: \text{G is a graph containing a clique of size $\ge$ k}\}$
\end{center}
Note that CLIQUE is decision problem whose proof certificate would be a set of vertices in $G$ on which the induced subgraph is claimed to be a clique\footnote{A clique is a complete subgraph.} of size at least k. Verifying that claim should be as easy as verifying that all vertices are valid, pairwise-adjacent in $G$ and that their count is at least k. All of that can be done in polynomial time making CLIQUE an NP problem. In fact, it was also proven to be NP-complete \cite{Kar72}. 
Now let's bring back the halting problem. HALT is surely not in NP because it is undecidable, hence undecidable by an NTM. Despite the apparent irrelevance of HALT to CLIQUE, we can, in fact, find a reduction from CLIQUE to HALT.
\begin{theorem}
    CLIQUE $\leq_{\mathcal{P}}$ HALT.
\end{theorem}
\begin{proof}

Let $M\langle G,k \rangle$ be a TM that on input graph $G$ and integer $k$: 
\begin{enumerate}
    \item generates all possible subsets of vertices in graph $G$.
    \item checks every generated subset for whether it forms a clique of size at least $k$.
    \item if such subset is found, it \emph{accepts}.
    \item otherwise, goes to step 1.
\end{enumerate}
\begin{remark}
If no satisfying subset exists, the machine $M$ effectively loops forever. As such, $M$ halts only when a satisfying clique exists. 
\end{remark}
Now we can establish the reduction from CLIQUE to HALT as: 
\begin{center}
    HALT$\langle M, \langle G,k \rangle\rangle$ $\iff$ $\langle G,k \rangle \in \text{CLIQUE}$
\end{center}
Note that $M$ is an EXPTIME Turing machine, but the reduction does not involve actually running it, therefore, CLIQUE is polynomially reducible to HALT.
\end{proof}
As such, we have effectively shown that HALT is, in fact, an NP-hard problem that is harder than NP-complete problems, refuting the common misconception that NP-complete problems are the hardest class of problems. That being said, they are most surely the hardest problems in the NP class. What also makes them interesting is that they form an equivalence class over polynomial-time reductions. That is, efficiently solving one of them is tantamount to efficiently solving all of them since we can merely reduce any other NP-complete problem (or any NP problem for that matter) to that problem we managed to efficiently solve.

\chapter{Boolean Satisfiability (SAT)}

In the field of theoretical complexity, SAT holds a unique place for several reasons including the fact that it has the simplest formulation of all other NP-complete problems, and it was the very first to be proven NP-complete along with 21 graph and combinatorial problems by \cite{Kar72}.

\begin{definition}
A Boolean formula $\phi$ is satisfiable if there exists an interpretation (i.e., variable assignment) that makes it hold true.
\end{definition}

\noindent As such, we can define the SAT language as:
\begin{center}
    $\text{SAT} = \{\phi \;|\: \phi \text{ is satisfiable}\}$
\end{center}

\begin{definition}
A Boolean formula $\phi$ is valid if it holds true for every possible interpretation (i.e., variable assignment).
\end{definition}

\begin{lemma}
There is a duality between validity and satisfiability:
\begin{enumerate}
  \item $\phi$ is satisfiable iff $\neg \phi$ is not valid.
  \item $\phi$ is valid iff $\neg \phi$ is unsatisfiable.
\end{enumerate}
\end{lemma}

In this chapter, we explore some of the most important and interesting aspects about the SAT problem, including its variations, and special properties.

\section{General SAT}
The language of SAT is classified under multiple categories depending on the form of its Boolean formulas. Some of these variations are easier than others, and many others are, in fact, equivalent despite their apparent dissimilarities. 
There are two major normal forms in which any Boolean formula can be expressed: Conjunctive Normal Form (CNF) and Disjunctive Normal Form (DNF). As the name suggests, CNF is a conjunction of disjunctions (i.e., disjunctive clauses) while DNF is a disjunction of conjunctions (i.e., conjunctive clauses). It should be easy to see that formulas in DNF are much easier to solve as one only needs to show that one member conjunctive clause is satisfiable for the whole disjunctive system to be so. Deciding whether a conjunction is satisfiable is as easy as checking that it does not contain a contradiction; that is, it does not simultaneously contain a variable and its negation. This check can be done efficiently in linear time. In Boolean algebra, any formula can expressed in CNF as well as DNF; however, the conversion process might be exponentially long in terms of the time it takes and the length of the resulting formula. For instance, consider the following example of a CNF formula\footnote{Note that every literal in a Boolean formula (e.g., $X_1$, $Y_1$, etc.) can either refer to a variable or its negation, and different literals do not necessarily refer to different variables.}:
\begin{center}
    $(X_1\lor Y_1)\land(X_2\lor Y_2)\land ... \land(X_n\lor Y_n)$
\end{center}

This formula can be converted to DNF by applying the distributive property of Boolean algebra which will result in an expression containing $2^n$ terms. If we interchange conjunctions and disjunctions in the above expression, we get the following DNF formula:
\begin{center}
    $(X_1\land Y_1)\lor(X_2\land Y_2)\lor ... \lor(X_n\land Y_n)$
\end{center}

Likewise, if we wish to convert that formula to an equivalent one in CNF, we get the following expression:

\begin{center}
    $(X_1\lor X_2 \lor ... \lor X_n)\land$  \\
    $(Y_1\lor X_2 \lor ... \lor X_n)\land$    \\
    $(X_1\lor Y_2 \lor ... \lor X_n)\land$    \\
    $(Y_1\lor Y_2 \lor ... \lor X_n)\land...\land$    \\
    $(Y_1\lor Y_2 \lor ... \lor Y_n)$
\end{center}
Each disjunctive clause above contains $n$ literals, and the whole expression contains every possible configuration of such clause on $X$s and $Y$s which amounts to a total of $2^n$ clauses. However, although converting an arbitrary Boolean formula $\phi$ into an equivalent one in CNF or DNF potentially takes exponential time, we actually do not need to find an \emph{equivalent} one to determine whether $\phi$ is satisfiable through solving CNF or DNF. Instead, we only need to construct an \emph{equisatisfiable} formula to $\phi$, that is, a formula $\phi'$ that is satisfiable if and only if $\phi$ is satisfiable. A simple example to distinguish between logical equivalence and equisatisfiability is the following two formulas:
\begin{center}
    let $\phi$ be a Boolean formula \\
    let $\Psi = \phi \land z$
\end{center}
Note that if $\phi$ is satisfiable, then $\Psi$ must also be satisfiable since we can use the same satisfying variable assignment of $\phi$ and append to it a \textit{true} assignment for $z$. Conversely, if $\phi$ is unsatisfiable, then no possible assignment can satisfy $\Psi$ since it would always be the case that $\Psi = \textit{false} \land z$ which is always \textit{false}. Hence $\phi$ and $\Psi$ are indeed equisatisfiable. However, strictly speaking, they are not equivalent because they have different variables, so their satisfying variable assignments would be different.

The good news is there are algorithms that can transform an arbitrary Boolean formula into an equisatisfiable one in CNF in polynomial time such as the one devised by Karp \cite{Kar72}. However, the bad news is that no similar polynomial transformation is known for DNF, which is the much easier one to solve. In fact, the existence of such reduction would imply $\text{P}=\text{NP}$.  What makes efficient CNF reductions useful is that we can now transform the SAT problem into CNF-SAT without making extra assumptions about its complexity class since the conversion is PTIME anyways, so rest assured; CNF-SAT is just as hard! That is why SAT is generally used to mean CNF-SAT.

\section{Feasible SATs}
We have shown a glimpse of how difficult SAT could be, yet the situation is not all that gloomy because despite the fact that no general SAT solver has been found to have a polynomial worst-case complexity, many practical SAT instances that appear in real life can be quickly solved using heuristics even in millions of variables. Furthermore, some SAT versions are, in fact, in P. This subsection is dedicated to presenting some of them.
\subsection{2-SAT}
One of the most common feasible SATs is \textbf{2-SAT} which is a special case of CNF-SAT where each clause contains at most 2 literals. This variant can be efficiently solved by first writing each clause in its \textbf{implicative normal form} that is as simple as writing each disjunction as an equivalent implication. For instance, $x \lor y$ can be written as $\bar{x} \Rightarrow y$ or $\bar{y} \Rightarrow x$. Now, since all clauses are in conjunction, we can just chain all implications together and see if we obtain an implicative contradiction in form of $x \Rightarrow ... \Rightarrow \bar{x}$ AND $\bar{x} \Rightarrow ... \Rightarrow x$, in which case the formula would be unsatisfiable. Performing the latter check has a linear-time solution by \cite{aspvall1979linear} through constructing a implication graph where implications between literals correspond to edges between vertices. The algorithms is based on graph theory concepts such as strongly connected components and topological sorting (for finding an assignment) which are defined as follows.

\begin{definition}
A directed graph $G$ is said to be \textbf{strongly connected} if there exists a path between any two of its vertices.
\end{definition}
\begin{remark}
A strongly connected component is a subgraph that is strongly connected.
\end{remark}

\begin{definition}
For a directed acyclic graph $G$, \textbf{topological sort} is a linear ordering of $V(G)$ in which vertex $u$ comes before vertex $v$ if there is directed edge $uv$ from $u$ to $v$.
\end{definition}

\begin{algorithm}[H]
\SetAlgoLined
\SetKwInOut{Input}{Input}
\Input{Boolean formula $\phi$}
\KwResult{YES if $\phi$ is satisfiable, NO otherwise}
Convert $\phi$ to implicative normal form $\gamma$. \\
Construct graph $G$ with all literals in $\gamma$. \\
\ForEach{$\alpha = (x \Rightarrow y) \in \gamma$}{
    $E(G) := E(G) + \{xy, \bar{y}\bar{x}\}$
}
Let $\mathcal{CC}:= \text{set of maximal strongly connected components of } G$ \\
\ForEach{$\mathcal{C} \in \mathcal{CC}$}{
    \ForEach{$u \in V(\mathcal{C})$}{
        \If{$\bar{u} \in V(\mathcal{C})$}{%
            \Return NO
      }
    }
}
\Return YES
 \caption{2-SAT Solver}
 \label{algo:2sat}
\end{algorithm}
As shown, the algorithm constructs the implication graph of $\phi$ and checks whether a literal and its negation exist on the same strongly connected component. For instance, consider the following formula:

\begin{example}
\label{example:2sat}
Let $\phi = (a\lor \neg b)\land(\neg a\lor b)\land(\neg a\lor \neg b)\land(a \lor\neg c)$. Writing $\phi$ in implicative normal form yields the following respective implications:  

{\centering
    $\neg a\Rightarrow \neg b, a\Rightarrow b, a\Rightarrow \neg b,\neg a \Rightarrow\neg c$ \\
    and equivalently, $b \Rightarrow a, \neg b\Rightarrow \neg a, b\Rightarrow \neg a,c \Rightarrow a$ \\
}
\noindent Constructing the implication graph accordingly gives us the following graph:
\begin{figure}[h]
    \centering
    \includegraphics[width=.45\textwidth]{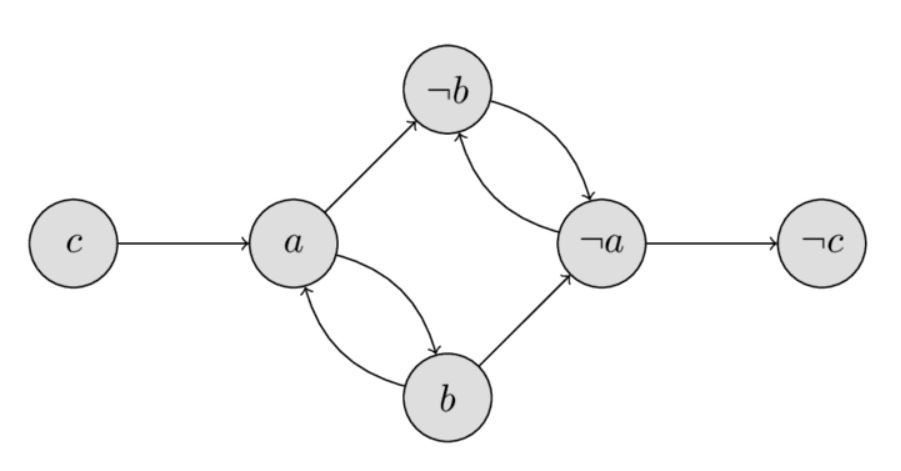}
    \caption{Example Implication Graph}
    \label{fig:imp_graph1}
\end{figure}
\begin{figure}[h]
    \centering
    \includegraphics[width=.45\textwidth]{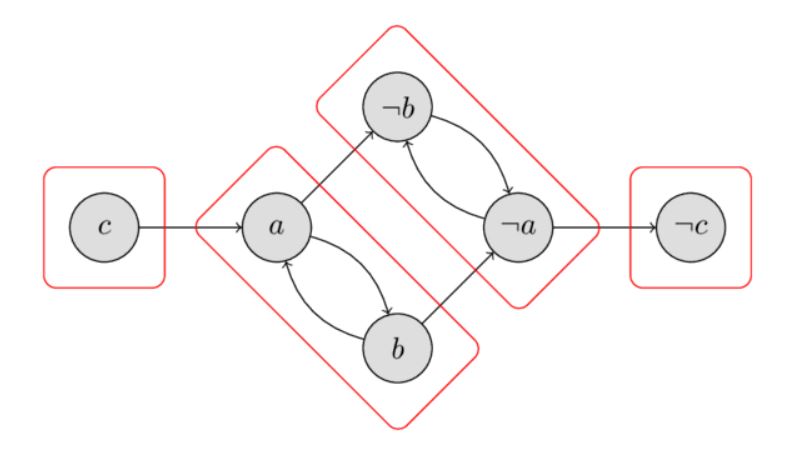}
    \caption{Implication Graph Strongly Connected Components}
    \label{fig:imp_graph2}
\end{figure}

\end{example}

As shown in Figure \ref{fig:imp_graph2}, the graph contains no contradicting literals on the same SCC, hence $\phi$ is statisfiable.
This algorithm \emph{decides} 2-satisfiability, but how about finding a satisfying variable assignment? Here comes an interesting property of some decision problems called \textbf{self-reducibility}. Generally, it refers to the ability of construct a reduction from a problem's decision version to its function version. For example, the decision version of SAT (which we are mostly dealing with) is deciding whether a formula is satisfiable whereas the function version is concerned with finding one such assignment. Luckily, 2-SAT is self-reducible. 
We can now use the 2-SAT solver to find a satisfying interpretation. One straightforward way to do so would be to iterate over each variable $x$ of the formula $\phi$, set it to \textit{true} in $\phi$ then test if the resulting formula is still satisfiable. If it is, then $x$ can be \textit{true} in the desired interpretation. Otherwise, if the resulting formula becomes unsatisfiable, then $x$ must be \textit{false}. Note that at every step, previous assignments need to be assumed in the substitution. Another more efficient way to find a satisfying assignment is through an extension to Algorithm \ref{algo:2sat} which sorts strongly connected components topologically where $\textit{comp}[u] \leq \textit{comp}[v]$ if there is a path from $u$ to $v$ where $\textit{comp}[u]$ denotes the topological order index of the SCC to which vertex $u$ belongs. Then, we assign variable $x$ as \textit{false} if $\textit{comp}[x] < \textit{comp}[\neg x]$, otherwise, we assign it as \textit{true}. We are not going to go into the proof of correctness for such method, but it to give some perspective, we can apply it to Example \ref{example:2sat}. From the marked implication graph, we can note the following relations: $\textit{comp}[a] < \textit{comp}[\neg a]; \textit{comp}[b] < \textit{comp}[\neg b]; \textit{comp}[c] < \textit{comp}[\neg c]$. As such, the corresponding truth assignment is $a=\textit{false},b=\textit{false},c=\textit{false}$ which indeed satisfies $\phi$.

\subsection{HORN-SAT}
Horn satisfiability is a variant of CNF-SAT where each clause is a Horn clause where at most one literal is positive (i.e. unnegated). Note that HORN-SAT places no restriction on the number of literals in each clause which might give the impression that it is intractable\footnote{Intractable problems are ones for which no efficient solution exists.}. However, despite the looks of it, HORN-SAT can be solved in polynomial time -- in linear time even. First, we introduce the method of \textbf{Unit propagation (UP)} or \textbf{Boolean Constraint propagation (BCP)}. It is a procedure by which CNF formulae can be simplified via exploiting unit clauses (i.e. clauses containing one literal).

\begin{algorithm}[H]
\SetAlgoLined
\SetKwInOut{Input}{Input}
\SetKwRepeat{Do}{do}{while}
\Input{Boolean formula $\phi$}
\KwResult{$\mathcal{U}(\phi): \text{the reduced formula of }\phi$}
Let $\mathcal{S}_u := \text{set of all unit clauses of } \phi$ \\

\While{$\mathcal{S}_u \neq \emptyset$} {
\ForEach{literal $x \in \mathcal{S}_u$}{
    Eliminate the unit clause of $x$ from $\phi$ \\
    \ForEach{clause $\gamma\ \in \phi $}{
        \uIf{$x \in \gamma$}{%
            Remove $\gamma$ from $\phi$
        }
        \uElseIf{$\bar{x} \in \gamma$}{
            Remove $\bar{x}$ from $\gamma$
        }
    }
}
Update $\mathcal{S}_u$ \\
}
 \caption{Unit Propagation (UP)}
\end{algorithm}
There are a few things to show about this method before we show how it can be used to efficiently solve HORN-SAT. 
\begin{proposition}
If $\Gamma$ is a Horn formula, then $\mathcal{U}(\Gamma)$ is also Horn.
\end{proposition}
\begin{proof}
It is easy to infer this from the UP procedure by noting that it either eliminates literals or entire clauses. It does not introduce new positive literals into clauses, so the Horn property should be preserved.
\end{proof}

\begin{lemma}
\label{lemma:horn_1_1}
If $\Gamma$ is an unsatisfiable formula, then $\Gamma$ contains at least one positive clause and one negative clause.
\end{lemma}
\begin{proof}
For the sake of contradiction, assume that $\Gamma$ has no positive clauses, then every clause of $\Gamma$ contains at least one negative literal. Thus, we can form a satisfying interpretation by assigning all variables to \textit{false}. Likewise, we can reach a contradiction if we assume $\Gamma$ has no negative clauses.
\end{proof}

\begin{theorem}
\label{them:horn-sat}
A Horn formula $\Gamma$ is satisfiable $\iff$ $\emptyset \notin \mathcal{U}(\Gamma).$ where $\emptyset$ is the empty clause.
\end{theorem}

\begin{proof}
$(\Rightarrow):$ if $\emptyset \in \mathcal{U}(\Gamma).$, then $\Gamma$ cannot be satisfiable because that implies that while eliminating some literal $x$, UP encountered a clause that only contains $\bar{x}$. That means that $x$ and $\bar{x}$ were in conjunction, so $\Gamma$ contains a contradiction.
$(\Leftarrow):$ If $\Gamma$ is Horn, then $\mathcal{U}(\Gamma).$ is also Horn. Now suppose $\mathcal{U}(\Gamma).$ is unsatisfiable. By Lemma \ref{lemma:horn_1_1}, $\mathcal{U}(\Gamma).$ contains at least one positive clause and one negative clause. A positive Horn clause must either be a unit clause or an empty clause. However, $\mathcal{U}(\Gamma).$ cannot contain a unit clause as it would have been eliminated during UP. Thus, the positive Horn clause must be the empty clause $\emptyset$.
\end{proof}
\noindent Therefore, HORN-SAT can be efficiently solved by performing UP then applying Theorem \ref{them:horn-sat}.

\section{Intractable SATs}
In this section, we show two famous intractable versions of SAT and their relationship to general SAT.
\subsection{3-SAT}
As the name suggests, 3-SAT is another special case of CNF-SAT where each clause has at most three literals. Knowing that 2-SAT is tractable in polynomial-time, one might be deceived into thinking that since 3-SAT seems slightly more constrained than 2-SAT, it must be only slightly harder. Nonetheless, that is far from reality because, as we shall shortly see, 3-SAT is in fact NP-complete, hence equivalent in hardness to general SAT itself. To show this, we present a polynomial reduction from SAT to 3-SAT which, along with the fact that 3-SAT is apparently in NP, proves that 3-SAT is NP-complete. 

\begin{theorem}
SAT $\leq_{\mathcal{P}}$ 3-SAT.
\end{theorem}
\begin{proof}
To prove this, we consider an arbitrary CNF-SAT formula $\phi$, and construct an equisatisfiable 3-SAT formula $\Psi$. We do this transformation for any clause $\mathcal{C} \in \phi$ by replacing it with a set $\mathcal{Z}$ of 3-CNF clauses. For any $\phi$ clause, there are four cases:

\begin{enumerate}[label=\textbf{Case \arabic*:}, leftmargin=4em]
    \item \textbf{$\mathcal{C}$ has 1 literal:} let $\mathcal{C}=a$ where $a$ is a literal. We introduce two new variables $z_1$ and $z_2$ to replace $\mathcal{C}$ by the following four 3-CNF clauses: 
    
    {\centering
        $(a\lor z_1 \lor z_2), (a\lor \neg z_1 \lor z_2), (a\lor z_1 \lor \neg z_2), (a\lor \neg z_1 \lor \neg z_2)$ \\
    }
    Note that the conjunction of all four clauses should resolve to $(a)$ since any truth assignment of $z_1$ and $z_2$ will satisfy exactly three of the four clauses leaving only one where the two disjunct $z$-literals evaluate to \textit{false} which yields $(a)$.
    
    \item \textbf{$\mathcal{C}$ has 2 literals:} let $\mathcal{C}=(a_1 \lor a_2)$. We introduce a new variable $z$ to replace $\mathcal{C}$ by the following two 3-CNF clauses:
    
    {\centering
        $(a_1 \lor a_2 \lor z), (a_1 \lor a_2 \lor \neg z)$ \\
    }
    Similarly, the conjunction of both clauses should resolve to $(a_1 \lor a_2)$, so the resulting expression is equisatisfiable to $\mathcal{C}$.
    \item \textbf{$\mathcal{C}$ has 3 literals:} leave $\mathcal{C}$ intact; it is already 3-CNF.
    \item \textbf{$\mathcal{C}$ has more than 3 literals:} let $\mathcal{C}=(a_1 \lor a_2 \land ... \land a_n)$ where $n > 3$. Here, we introduce $n-3$ new variables $z_1, z_2, ..., z_{n-3}$ replacing $\mathcal{C}$ by the following $n-3$ clauses:
    
    {\centering
        $(a_1 \lor a_2 \lor z_1), (\neg z_1 \lor a_3 \lor z_2), ... ,(\neg z_i \lor a_{i-2} \lor z_{i+1}), (\neg z_{i+1} \lor a_{i-1} \lor z_{i+2}), ..., (\neg z_{n-3} \lor a_{n-1} \lor a_n)$ \\
    }
    
    Unlike the first two cases, the conjunction of these $\mathcal{Z}$ clauses is not equivalent to $\mathcal{C}$; however, it is equisatisfiable to it. We can demonstrate that equisatisfiability by cases as follows: \\
    \textbf{Case 4a:} $\mathcal{C}$ is satisfiable: \\
    The satisfiability of $\mathcal{C}$ ensures that at least one of $a_1, a_2,..,a_n$ must be \textit{true} in any satisfying assignment of $\phi$. Using that fact, we can come up with a satisfying assignment for all $\mathcal{Z}$ clauses based on the location of that \textit{true} literal:
    \begin{itemize}
        \item If any of $a_1,a_2$ is \textit{true}, then set all $z$-variables to \textit{false}. The truth of $a_1 \lor a_2$ satisfies the first clause while the \textit{false}hood of all $z$-variables satisfies all other clauses since every $\mathcal{Z}$ clause after the first one contains a negated $z$-variable. Thus, all $\mathcal{Z}$ clauses are satisfiable.
        
        \item If $a_i$ is \textit{true} where $3 \leq i \leq n$, then set $z_1$ to \textit{true} and $z_2,...,z_{n-3}$ to \textit{false}. Similarly, the truth of $z_1$ satisfies the first clause while the \textit{false}hood of all other $z$-variables satisfies all other clauses since every other $\mathcal{Z}$ clause after the first one contains a negated $z$-variable. Thus, all $\mathcal{Z}$ clauses are satisfiable.
    \end{itemize}
    
    \textbf{Case 4b:} $\mathcal{C}$ is unsatisfiable: \\
    The unsatisfiability of $\mathcal{C}$ ensures that none of $a$-literals can be \textit{true} without falsifying some other clause in $\phi$. That means that we need to assume that all $a$-literals must be \textit{false}. Having all $a$-literals be \textit{false} reduces our $\mathcal{Z}$ clauses into:
    
    {\centering
        $z_1, (\neg z_1 \lor z_2), ... ,(\neg z_i \lor z_{i+1}), (\neg z_{i+1} \lor  z_{i+2}), ..., \neg z_{n-3}$ \\
    }
    Clearly, to have all $\mathcal{Z}$ clauses \textit{true}, then $z_1$ must be \textit{true} and $z_{n-3}$ must be \textit{false} since they form two unit clauses. Notice how if we  keep doing this substitution for the first and last literals (the ones forming unit clauses) accordingly, the truth of the first variable will falsify the first literal in the next intermediate clause. Similarly, the \textit{false}hood of the last literal removes the second literal in the previous clause. For example, consider the following reduced $\mathcal{Z}$ clauses: 
    
    \begin{center}
        $z_1, (\neg z_1 \lor z_2),(\neg z_2 \lor z_3),(\neg z_3 \lor z_4), \neg z_4$ \\
        $\downarrow$ \\
        $z_2,(\neg z_2 \lor z_3),\neg z_3$ \\
        $\downarrow$ \\
        $z_2,\neg z_2$ \\
        $\downarrow$ \\
        (contradiction)
    \end{center}
    
    As such, the conjunction of all $\mathcal{Z}$ clauses ultimately result in a contradiction, thereby making them unsatisfiable.
    
\end{enumerate}
Therefore, $\mathcal{C} \in \text{SAT} \iff \left(\bigwedge\limits_{\gamma \in \mathcal{Z}} \gamma \right) \in \text{SAT}$, i.e. they are equisatisfiable.

Finally, this reduction is evidently polynomial since we only add either a constant number of new clauses/variables (cases 1, 2, \& 3) or a polynomial number thereof (case 4 with $n-3$ clauses). 

\end{proof}

Thus, we have shown that 3-SAT is indeed NP-complete.

\subsection{MAX-SAT}
Despite it being listed as an SAT variant, \textbf{Maximum Satisfiability} is in fact more general than general SAT as it asks about the maximum number of clauses in a formula that are satisfiable with one variable assignment. The decision version of that asks whether given a Boolean formula $\phi$ in CNF, there exists an interpretation that satisfies at least $k$ clauses of $\phi$. 

\begin{definition}
    $\text{MAX-SAT} = \{\langle \phi,k \rangle \;|\: \text{formula $\phi$ has at least $k$ jointly satisfiable clauses}\}$
\end{definition}

Now it seems clear why this goal is more general than general SAT because one can obtain an answer to the latter by merely replacing $k$ with the total number of clauses in $\phi$. As such, SAT easily reduces to MAX-SAT, making it NP-hard. Moreover, SAT and MAX-SAT share the same type of certificates being variable assignments, so verifying MAX-SAT solutions is as simple as substituting these assignments and counting the clauses they make \textit{true}. That means that the decision problem of MAX-SAT is indeed in NP, hence NP-complete.

\begin{example}
Consider the following Boolean formula: 
    $\phi = (a \lor b)\land(a \lor \neg b)\land(\neg a \lor b)\land(\neg a \lor \neg b)$. 
$\phi$ is evidently unsatisfiable on all possible truth assignments of $a$ and $b$. However, any truth assignment will satisfy three out of all four clauses, so $\langle \phi,3 \rangle \in \text{MAX-SAT}$.
\end{example}

\noindent Unsatisfied clauses in $\phi$ can be viewed as a conflict with other satisfied ones. In that sense, maximum satisfiability aims to minimize conflicting clauses in a given formula. That is why MAX-SAT has many useful applications in the field of software. One example of that is software package upgradeability where we have a set of packages each requiring other packages while also being in conflict with others. In that scenario, we want to find the maximum number of packages that we can install without causing any conflicts, which is a standard application of MAX-SAT since packages can be easily modelled as Boolean variables. A more complex application would be error localization in C code where we have a faulty program and we aim to find the smallest segment of code that is causing that fault. This is quite challenging since errors in some lines can propagate to other lines, giving the illusion that those other errors are erroneously written while they might in fact be correct. Luckily, there are numerous heuristic approximations around MAX-SAT that are very good at solving real-life instances of the problem, albeit not any general arbitrary case.

\newpage
\section{The Cook-Levin Theorem (CLT)}
Now we arrive at one of the most central theorems in theoretical complexity which was arguably the last notable stride on the quest for answering the P vs. NP dilemma. The reason why it is of great significance is that it simplifies the question of finding whether the two classes of problems are equivalent to merely determining whether we can efficiently solve a single NP-complete problem. In spite of that reduction, the problem remains unsolved for nearly half a century! 
The theorem was named after Stephen Cook and Leonid Levin who both reached it independently. 
\begin{theorem}
SAT is NP-complete.
\end{theorem}
The following proof of CLT is based on the works of \cite{hartmanis1982computers, sipser2012introduction} and it uses a few concepts that were introduced earlier in this paper.

\begin{tcolorbox}[title=Proof Idea]
To prove that a problem is NP-complete, we need to first show that it is in NP, then we show that every problem in NP is polynomially reducible to it. It is easy to show that SAT is in NP which will be presented shortly. The second part is a proof by construction in which we need to come up with a polynomial many-to-one reduction to SAT that works for any problem in NP. We construct a Boolean formula $\Psi$ that is satisfiable if and only if a TM for some NP problem $L$ accepts on some input $I$. More formally, $I \in L \iff \phi \in \text{SAT}$, which is the essence of reduction.
\end{tcolorbox}

\begin{proof}
To show that SAT is in NP, it suffices to see that it can be solved in polynomial time by an NDTM that guesses all possible variable assignments at every branch and accepts when it finds a satisfying assignment. 
Now we move to constructing a polynomial many-to-one from an arbitrary problem $L$ in NP to SAT. By definition, there exists a non-deterministic Turing machine $N_L$ that decides $L$ in $p(n)$ steps where $p$ is a polynomial and $n$ is the size of input $I$. Here, we introduce an auxiliary construct to help us describe $N_L$'s computation in a tabular form called \textbf{tableau}. As shown in Figure \ref{fig:tableau}, every row contains one branch configuration on input $I$ starting with the initial configuration. The row format is set up so that every configuration is bounded by two \# symbols whose relevance will later be shown. The configuration itself is an array of cells containing configuration state $q \in Q$ along with tape contents $T_i \in \Gamma$ -- using the same Turing machine set denotations in section \ref{section:ord_tm}.
\begin{figure}[h]
    \centering
    \includegraphics[width=.65\textwidth]{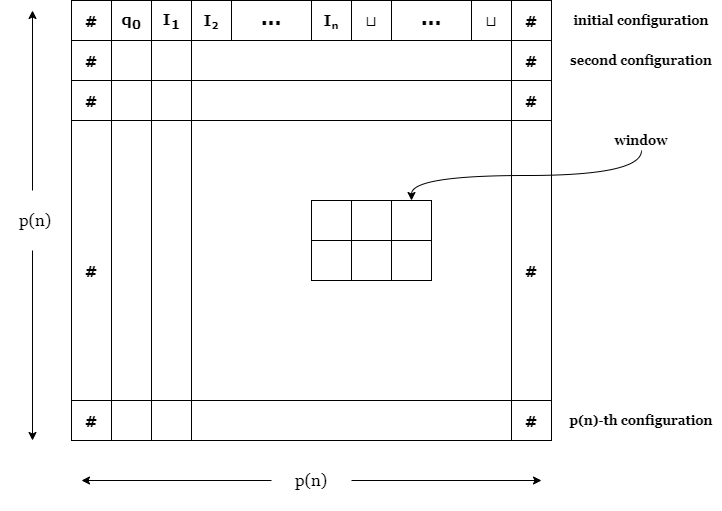}
    \caption{$p(n)\times p(n)$ configurations tableau}
    \label{fig:tableau}
\end{figure}

The state cell appears right before the symbol to which the tape head is pointing. That way, it represents both the state and the tape head location. That is why it appears before the tape sub-array in the first row since the tape head initially points to the first symbol on the tape.
Furthermore, rows (i.e. configurations) are ordered by their occurrence in the computational branch, that is, every configuration row is the result of applying the transition function $\delta$ on its predecessor. A \textbf{window} is $2 \times 3$ sub-matrix of the tableau, which will be used later to ensure the validity of configurations. It is easy to see why the tableau has $p(n)$ rows because every configuration corresponds to one step. Though, a tableau is not required to be exactly $p(n)$ columns wide, but we know that $N_L$ cannot consider more than $p(n)$ cells because that would take more than $p(n)$ contradicting the definition of $p$. Accordingly, tableau width was set as $p(n)$ containing only considered cells. In the case of a configuration requiring a smaller width to represent, its tape cells are padded with the blank symbol $\sqcup$. 

We say that a computation tableau is \textit{accepting} if it contains at least one accepting configuration row. More formally, a tableau if it satisfies the condition $\exists i X_{i,2,q_{accept}}$.

For convenience, we list new notations used in this proof along with their intended meaning in the following table:

\begin{table}[h!]
\label{table:clt-symbols}
\centering
\bgroup
\def\arraystretch{1.3}
 \begin{tabular}{||c | c||} 
 \hline
 \textbf{Symbol} & \textbf{Interpretation} \\ [0.5ex] 
 \hline\hline
 $cell[i, j]$ & Symbol written on the cell of row $i$ and column $j$ \\ \hline
 $X_{i,j,s}$ & $cell[i, j] = s$ \\ \hline
 $\mathcal{C}$ & $Q \cup \Gamma \cup \{\#\}$ \\ \hline
 $W_{i,j}$ & The $2\times 3$ window whose  top-left corner is $cell[i,j]$ \\ \hline
 $\mathcal{L}(W_{i,j})$ & Window $W_{i,j}$ is legal \\ [0.5ex] \hline
 \end{tabular}
 \egroup
 \caption{Used Symbols Interpretations}
\end{table}

Next, we start defining a Boolean formula $\Psi$ whose satisfiability corresponds to the existence of at least one \emph{accepting} tableau for $N_L$ on input $I$. Remember that a tableau is \textbf{one} computational branch of potentially many other branches on the computational tree of $N_L$. Thus, we only need to worry about the existence of an accepting branch. For convenience and generality, $\Psi$ will be expressed in terms of variables in the form $X_{i,j,s}$.

To achieve that, we need to map out the exact constraints that ensure that a tableau is valid, that is, it corresponds to valid computational branch. In this proof, we group those constraints into four sub-formulas whose conjunction constitutes $\Psi$. They are as follows:

\begin{table}[h!]
\label{table:clt-subformulas}
\centering
\bgroup
\def\arraystretch{1.3}
 \begin{tabular}{||c | c||} 
 \hline
 \textbf{Sub-formula} & \textbf{Constraint} \\ [0.5ex] 
 \hline\hline
 $\Psi_{cell}$ & Every cell has exactly one defined symbol on it \\ \hline
 $\Psi_{initial}$ & The first row contains a valid initial configuration \\ \hline
 $\Psi_{accept}$ & The tableau contains an accepting configuration \\ \hline
 $\Psi_{move}$ & Each configuration row is a legal transition from the previous one \\ [0.5ex] \hline
 \end{tabular}
\egroup
 \caption{$\Psi$ Sub-formulas and Their Interpretation}
\end{table}

On that basis, if we manage to capture all four constraints using Boolean variables, then we guarantee a correspondence between a valid computation of $N_L$ and a variable assignment on $\Psi$:

{\centering
$\Psi = \Psi_{cell}\land\Psi_{initial}\land\Psi_{accept}\land\Psi_{move}$ \\
}

The first three sub-formulas are relatively straightforwards while $\Psi_{move}$ might seem relatively more involved. To begin with, we can define $\Psi_{cell}$ as:

\[ \Psi_{cell} = \bigwedge\limits_{1 \leq i, j \leq p(n)} \left[\left(\bigvee\limits_{s \in \mathcal{C}} X_{i,j,s}\right) \land \left(\bigwedge\limits_{\substack{s, r \in \mathcal{C} \\ s \neq r}} \neg\left(X_{i,j,s} \land X_{i,j,r}\right)\right)\right] \]

The expression above translates to the property that every cell in the tableau needs to first have a symbol from the given list of allowed symbol (i.e. $\mathcal{C}$) AND not have two different symbols simultaneously.

Next, we define $\Psi_{initial}$ as:

\[ \Psi_{initial} = X_{1,1,\#}\land X_{1,2,q_0}\land \left(\bigwedge\limits_{\substack{3 \leq j \leq n + 2 \\ s \in I}} X_{1,j,s}\right) \land \left(\bigwedge\limits_{n+3 \leq j \leq p(n)-1} X_{1,j,\sqcup}\right) \land X_{1,p(n),\#} \] 

Looking at the expression for $\Psi_{initial}$, one can see the clear resemblance between it and the first row indicated in Figure \ref{fig:tableau} as it is describing exactly the same format for the initial configuration row (i.e. the 1st one). 

Moving on to defining $\Psi_{accept}$ which is arguably the most straightforward:

\[ \Psi_{accept} = \bigvee\limits_{1 \leq i,j \leq p(n)} X_{i,j, q_{accept}} \]

This expression essentially ensures that there is an accepting state cell somewhere in the tableau; otherwise, it would stand for either a \textit{rejecting} or an incomplete computational branch.

Finally, we set the stage for defining the $\Psi_{move}$ sub-formula to ensure the legality of all transitions. First, we can see that a transition is represented by two consecutive rows since every row (except for the first one) follows from the previous one. However, it would be infeasible to come up with a closed form for a valid transition between two complete rows. Here, windows along with the \# delimiters come into play. Using windows, we would only need to formally describe the validity of any $2\times 3$ sub-matrices which is considerably easier to accomplish. 

A window is \textbf{legal} if it can follow from a valid transition function $\delta$ for $N_L$. Note that this local notion of \textbf{legality} does in fact guarantee that no two windows contradict each other over the entire tableau due to the non-determinism of $N_L$'s transition function allowing it to take multiple different actions from the same configuration.

To illustrate, consider an example of a non-deterministic $\delta$ function that recognizes two symbols $\{a,b\}$does the following based on three states $q_1, q_2,q_3$:
\begin{itemize}
    \item In $q_1$, invert the currently read symbol (e.g., from $a$ to $b$), transition to state $q_2$ and move right.
    \item In $q_2$, preserve the currently read symbol (e.g., $a$ stays $a$), transition to state $q_3$ and move left.
    \item In $q_3$, non-deterministically either:
    \begin{itemize}
        \item Preserve the currently read symbol, transition to state $q_1$ and move right.
        \item Invert the currently read symbol, stay in $q_3$ and move left.
    \end{itemize}
\end{itemize}

Formally, we can express this $\delta$ function using the following table:

\begin{table}[h!]
\label{table:delta_func_clt}
\centering
\bgroup
\def\arraystretch{1.3}
 \begin{tabular}{||c | c||} 
 \hline
 $Q\times\Gamma$ & $\delta(Q\times\Gamma)$ \\ [0.5ex] 
 \hline\hline
 ($q_1$, $a$) & $\{(q_2, b, R)\}$ \\ \hline
 ($q_1$, $b$) & $\{(q_2, a, R)\}$ \\ \hline
 ($q_2$, $a$) & $\{(q_3, a, L)\}$ \\ \hline
 ($q_2$, $b$) & $\{(q_3, b, L)\}$ \\ \hline
 ($q_3$, $a$) & $\{(q_1, a, R), (q_3, b, L)\}$ \\ \hline
 ($q_3$, $b$) & $\{(q_1, b, R), (q_3, a, L)\}$ \\ [0.5ex] \hline
 \end{tabular}
 \egroup
 \caption{Example $\delta$ Mapping}
\end{table}

Now let's see a few examples and non-examples of \textbf{legal} windows in light of that $\delta$ function:

\begin{table}[h!]
\label{fig:legal_windows}
\centering
\bgroup
\def\arraystretch{1.1}

\begin{tabular}{ P{0.1\textwidth} @{}c@{} P{0.1\textwidth} @{}c@{} P{0.1\textwidth} @{}c@{} }

(a) & 
\begin{tabular}{|c|c|c|} 
\hline
$b$ & $a$ & $b$ \\ \hline 
$b$ & $a$ & $b$ \\ \hline
\end{tabular} & 
(b) & 
\begin{tabular}{|c|c|c|} 
\hline
$b$ & $q_1$ & $b$ \\ \hline 
$b$ & $a$ & $q_2$ \\ \hline
\end{tabular} & 
(c) & 
\begin{tabular}{|c|c|c|} 
\hline
$b$ & $q_3$ & $a$ \\ \hline 
$a$ & $a$ & $q_1$ \\ \hline
\end{tabular} \\ [5ex]

(d) & 
\begin{tabular}{|c|c|c|} 
\hline
$b$ & $a$ & $a$ \\ \hline 
$a$ & $a$ & $a$ \\ \hline
\end{tabular} & 
(e) & 
\begin{tabular}{|c|c|c|} 
\hline
$\#$ & $a$ & $q_1$ \\ \hline 
$\#$ & $a$ & $b$ \\ \hline
\end{tabular} & 
(f) & 
\begin{tabular}{|c|c|c|} 
\hline
$b$ & $q_3$ & $a$ \\ \hline 
$q_3$ & $b$ & $a$ \\ \hline
\end{tabular} \\

\end{tabular}
\egroup
\caption{Examples of Legal Windows}
\end{table}

In Table \ref{fig:legal_windows}, (a) is legal because we do not know the location of the tape head, so it might not be pointing at any cells in the first row. Window (d) is legal because the tape head might be pointing at the leftmost cell (but we would not be able to see it) while in state $q_3$ in which case, it can invert the $b$ to $a$ and move to the left. Windows (b) \& (e) are legal because they follow the correct $\delta$ rules. Note that in (c), the tape head takes a different decision from the one in (f) even though they are both in the same state $q_3$ and reading the same symbol $a$. That is still legal because the non-deterministic $\delta$ is allowed to make either of them.

\begin{table}[h!]
\label{fig:illegal_windows}
\centering
\bgroup
\def\arraystretch{1.1}

\begin{tabular}{ P{0.1\textwidth} @{}c@{} P{0.1\textwidth} @{}c@{} P{0.1\textwidth} @{}c@{} }

(g) & 
\begin{tabular}{|c|c|c|} 
\hline
$b$ & $a$ & $b$ \\ \hline 
$b$ & $b$ & $b$ \\ \hline
\end{tabular} & 
(h) & 
\begin{tabular}{|c|c|c|} 
\hline
$b$ & $a$ & $b$ \\ \hline 
$q_1$ & $a$ & $q_2$ \\ \hline
\end{tabular} & 
(i) & 
\begin{tabular}{|c|c|c|} 
\hline
$q_1$ & $a$ & $b$ \\ \hline 
$a$ & $q_2$ & $a$ \\ \hline
\end{tabular} \\ 

\end{tabular}
\egroup
\caption{Examples of Illegal Windows}
\end{table}

Table \ref{fig:illegal_windows} shows some examples that cannot result from the given $\delta$ function. In fact, the first two can never follow from any $\delta$ function whatsoever. Window (g) is illegal because the top middle symbol cell (containing $a$) was modified while it was not pointed to by the tape head. Window (h) is also illegal because it contains two state cells which cannot happen in this setting where we have only one tape head. On the other hand, window (i) is illegal because it simply violates the chosen $\delta$ rules as it preserve the symbol while in state $q_1$. Note that window (i), unlike other (g) and (h), is not absolutely illegal, that is, it can be a legal window for another $\delta$ function while (g) and (h) cannot.

Having explained the legality of windows, we need to show how it can be used to ensure the legality of row transitions. Towards that, we put forth the following proposition:

\begin{proposition}
\label{prop:r_r}
For any two consecutive configuration rows $\mathcal{R}_i$ and $\mathcal{R}_{i+1}$, if all windows sliding over both rows are legal, then $\mathcal{R}_{i+1}$ legally follows from $\mathcal{R}_i$. That is, $\mathcal{R}_i \text{ is legal} \Rightarrow \mathcal{R}_{i+1} \text{ is legal}$.
\end{proposition}
\begin{proof}
We can prove that proposition as follows. If $\mathcal{R}_i$ is illegal, then the proposition trivially follows. If $\mathcal{R}_i$ is legal, we consider two types of cells. Cells in $\mathcal{R}_i$ that are neither a state nor adjacent to a state cell will appear in the top-middle cell in some window\footnote{This is due to the row-bounding \# symbols.}. Since all windows on those two rows are legal, then that cell should appear unchanged in $\mathcal{R}_{i+1}$ in the same relative location where it is supposed to be in a legal transition. 

On the other hand, state cells in $\mathcal{R}_i$ must appear in the top-middle cell for some window, so the bottom three cells will be updated based on the $\delta$ function. That ensures that state cells along with state-adjacent cells are both correctly placed in $\mathcal{R}_{i+1}$. Therefore, if all windows of both rows are legal, then $\mathcal{R}_{i+1}$ must be legal, which concludes the proof of Proposition \ref{prop:r_r}.
\end{proof}

Having shown how window legality guarantees the legality of configuration row transitions, we can now define $\Psi_{move}$ as follows:

\[ \Psi_{move} = \bigwedge\limits_{\substack{1 \leq i \leq n - 2 \\ 1 \leq j \leq n - 3}} \mathcal{L}(W_{i,j}) \]

Note that we have not defined $\mathcal{L}(W_{i,j})$ in rigorous detail as it would take us far afield to precisely formulate it. However, such rigorous definition is not essential for this proof as long as $\mathcal{L}(W)$ can evidently be defined only in terms of $X_{i,j}$ on all of its six content cells. This is for two reasons. Firstly, only the cells of a window completely determine whether it is legal or not with respect to the given $\delta$ function. Secondly, the total number of distinct windows is finite since each window has a finite number of cells (six cells) and each cell can only take a finite number of values for any computation tableau on a given language\footnote{Note that $\mathcal{C}$ is finite since $Q$ and $\Gamma$ are finite.}.

Last but not least, it remains to show that this reduction from $L$ to SAT is polynomial-time by examining the size of $\Psi$ in terms of $p(n)$. Recall that $\Psi$ is entirely composed of variables in the form $X_{i,j,s}$ which amount to $\mathcal{O}(p(n)^2\times |\mathcal{C}|)$ variables. But since $|\mathcal{C}|$ is a constant defined by the machine $N_L$ (not input size $n$), then $\Psi$ has $\mathcal{O}(p(n)^2)$ variables. Next, we need to examine the size of each sub-formula of $\Psi$. Looking at $\Psi_{cell}$, it is clear that comprises $\mathcal{O}(p(n)^2\times |\mathcal{C}|^2) = \mathcal{O}(p(n)^2)$ literals. $\Psi_{cell}$ is a linear scan on the first row, so it involves $\mathcal{O}(p(n))$ literals. $\Psi_{accept}$ checks all tableau cells, so it has a size of $\mathcal{O}(p(n)^2)$. $\Psi_{move}$ iterates over all windows of which there are $\mathcal{O}(p(n)^2)$. The predicate $\mathcal{L}$, on any window $W_{i,j}$, has a fixed size irrespective of the input size $n$, so $\Psi_{move}$ is of size $\mathcal{O}(p(n)^2)$. Therefore, $\Psi$ has a polynomial size of $\mathcal{O}(p(n)^2)$. Note that this reduction merely writes out $\Psi$, so if $\Psi$ is of polynomial size, then the reduction would be of polynomial-time. As such, we have concluded the proof for the Cook-Levin theorem, showing that SAT is indeed NP-complete.

\end{proof}

\chapter{SAT Graph Reductions}
In this chapter, we present some of the most famous reductions from 3-SAT to classic NP-complete graph problems, namely, Clique, Hamiltonian Cycle, and 3-Coloring. All three problems can be easily shown to be in NP. As such, by establishing reductions from the NP-complete 3-SAT to these problems, we essentially prove their NP-completeness.

\section{3-SAT to CLIQUE}
Cliques are complete subgraphs, that is, a fully connected subset of vertices in some graph. The problem of finding cliques with some properties (e.g., size) is a very relevant one in graph theory. It is essentially a sibling problem of the Independent Set problem where we look for a pair-wise disjoint subset of vertices on some graph. CLIQUE is in NP since a proof certificate for it would consist of a set $\{v_1,...,v_k\}$ of vertices claimed to form a clique which can be checked in linear time. Next, we show that CLIQUE is intractable by reducing 3-SAT to it.

\begin{theorem}
3-SAT $\leq_{\mathcal{P}}$ CLIQUE.
\end{theorem}

\begin{proof}
To establish the reduction, for any 3-SAT Boolean formula $\Psi$ of $k$ clauses in variables $\{x_1,...,x_n\}$, we construct a graph $G$ such that $G$ contains a $k$-clique if and only if $\Psi$ is satisfiable. First, we allocate $k$ groups of vertices, one for each clause, that each contain three vertices corresponding to a literal in the respective clause. More formally, $V(G)=\{x_{i,j,h}\;|\: x_i \in \text{clause }j; \; h \in \{+,-\}\}$. The sign in the last index represents whether the literal is positive (i.e., unnegated variable) or negative (i.e., negated variable). Now, we need to form the edge set for $G$ which we do as follows:
\[E(G) = \{ \{x_{i_1,j_1,h_1}, x_{i_2,j_2,h_2}\}\;|\: (j_1 \neq j_2)\land(i_1=i_2 \rightarrow h_1=h_2) \}\]

Essentially, what this means is that all inter-group edges are present except for vertices of the same variable and opposite signs. This guarantees that any two connected vertices correspond to two (not necessarily different) literals that can be set to \textit{true} simultaneously. As such, a $k$-clique in $G$ corresponds to a valid truth assignment that satisfies all $k$ clauses since it satisfies exactly one literal in each clause. Note that this assignment might not be complete (i.e., does not include all variables) in case of repeated literals on the same clique, but that makes no difference since we can assign arbitrary truth values to the remaining variables without violating that assignment. This essentially proves the forward direction of the reduction. 

\begin{example}
Let $\Psi = (x_1\lor x_2 \lor x_3)\land (x_1\lor \neg x_2 \lor x_3)\land (\neg x_1\lor x_2 \lor \neg x_3)$. For such formula, the reduction graph $G$ looks as follows in Figure \ref{fig:3-clique}. For simplicity, the group and sign indices are omitted and replaced with the variable names colored with a different color for each group. Note the highlighted clique maps to the assignment that sets all variables to \textit{true} which indeed satisfies the given $\Psi$ formula.

\begin{figure}[h]
    \centering
    \includegraphics[width=.35\textwidth]{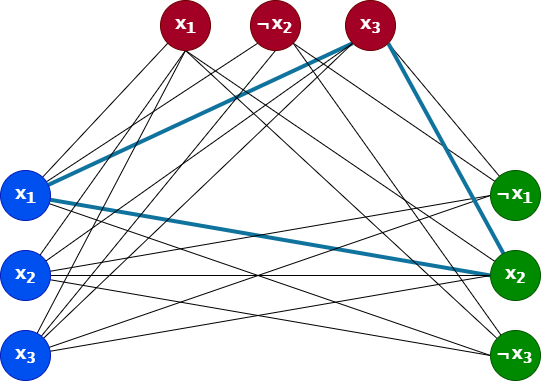}
    \caption{Example $\Psi$ Clique Reduction Graph}
    \label{fig:3-clique}
\end{figure}

\end{example}

To show the reverse direction, suppose $\Psi$ is satisfiable. That means that there exists a truth assignment $f: \{x_1,...,x_n\} \rightarrow \{\text{\textit{true}, \textit{false}}\}$ that satisfies at least one literal in every clause. Hence, their corresponding vertices must be fully connected (i.e., form a clique) since they belong to different clauses and they can evidently be set \textit{true} simultaneously. Thus, $G$ must contain a $k$-clique. That concludes the proof of reduction correctness.

Finally, we need to show that this reduction is polynomial-time. This is \textit{true} because $|G|= 3|\Psi|=3k$ where $|\Psi|$ is the number of clauses in $\Psi$. Note that this implies the size of $G$ (i.e., number of edges) is always $\mathcal{O}(\frac{n(n-1)}{2})=\mathcal{O}(n^2)$ where $n$ is the graph order, so $|E(G)|$ would also be polynomial in formula size $|\Psi|$. Therefore, this reduction indeed takes polynomial-time since constructing this graph takes polynomial-time.
\end{proof}

\section{3-SAT to HAM-CYCLE}
The Hamiltonian Cycle problem (HAM-CYCLE) is one of the original twenty-one NP-complete problems by Karp in 1971. A Hamiltonian cycle is a cycle that visits all vertices in a graph. In essence, HAM-CYCLE is a special case of the Travelling Salesman Problem (TSP) where we ask whether a graph contains a path that visits every vertex exactly once and returns to the starting vertex under a certain distance. Such path is also called a cycle and its length is the sum of the weights of all edges it traverses. As such, HAM-CYCLE is an instance of TSP where all edges have a weight equal to one, and the maximum path length is $|G|$ (number of vertices).

\begin{theorem}
3-SAT $\leq_{\mathcal{P}}$ HAM-CYCLE.
\end{theorem}

\begin{proof}
To establish the reduction, for any 3-SAT Boolean formula $\Psi$ of $k$ clauses in variables $\{x_1,...,x_n\}$, we construct a graph $G$ such that $G$ contains a Hamiltonian cycle if and only if $\Psi$ is satisfiable.
To achieve this, we need to encode truth assignments in the graph $G$ such that every truth assignment encodes a different path, and satisfying assignments encode a Hamiltonian one.
To encode assignments, we construct $n$ paths of length $2k$ where each path corresponds to a truth assignment on one variable. Unlike other graph reductions in this chapter, this reduction requires $G$ to be a directed graph for reasons that will follow shortly. Note that arrow-less edges indicate a bidirectional edge, that is, two opposite directed edges.

\begin{example}
Consider $\Psi = (x_1 \lor x_2 \lor x_3)\land(\neg x_2 \lor x_3 \lor \neg x_4)\land(\neg x_1 \lor x_2 \lor x_4)$. For simplicity, vertices of each variable path are numbered from $1$ to $2k$, colored differently. 
\end{example}

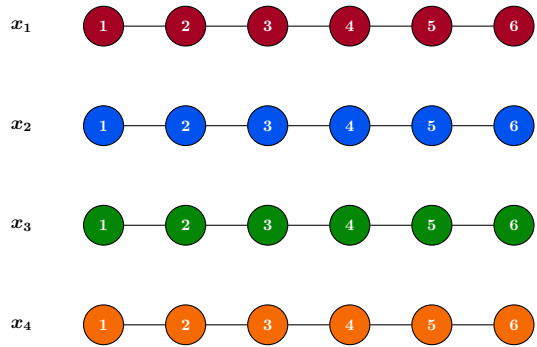
\begin{figure}[!h]
\centering

\begin{tikzpicture}[main/.style = {draw, circle, minimum size=2.5em}, scale=0.6, every node/.append style={transform shape}, 
        x1/.style={fill=dred, text=white},
		x2/.style={fill=dblue, text=white},
		x3/.style={fill=dgreen, text=white},
		x4/.style={fill=dorange, text=white}] 
		
\pgfmathsetmacro{\xsp}{1.8}
\pgfmathsetmacro{\ysp}{2.2}

\foreach \i in {1,...,4}
{
    \pgfmathsetmacro{\y}{-1*(\i - 1)*\ysp};
    \node[] (\i * 7) at (0,\y) {\large $\boldsymbol{x_{\i}}$};
    \foreach \j in {1,...,6}
    {
        \pgfmathtruncatemacro{\label}{\i*7 + \j};
        \pgfmathtruncatemacro{\prev}{(\label - 1)};
        \pgfmathsetmacro{\x}{\j * \xsp};
        \node[main,x\i] (\label) at (\x,\y) {\textbf{\j}};
        \ifthenelse{\j > 1}
        {\draw[-] (\label) -- (\prev);}
           
    }
}

\end{tikzpicture} 
\caption{Example $\Psi$ Assignment Graph}
\label{fig:ham-cycle}
\end{figure}

Under this scheme, traversing path $x_i$ from left to right (from $1\rightarrow 2k$) corresponds to assigning variable $x_i$ \textit{true} while traversing it from right to left (from $2k\rightarrow 1$) corresponds to an assignment of \textit{false}. This encodes a way to assign variables, but not one to make complete truth assignments on all variables using paths because the paths are not connected. To solve this, we connect them end-to-end in sequence as shown in Figure \ref{fig:ham-cycle1} so that a Hamiltonian path encodes a full truth assignment. Note that any Hamiltonian path must traverse every sub-path $x_i$ in only one direction, otherwise it would revisit vertices. That ensures that every Hamiltonian path in $G$ encode a \textbf{valid} truth assignment.

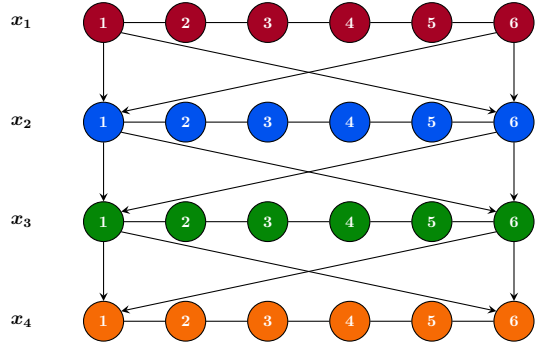
\begin{figure}[!h]
\centering

\begin{tikzpicture}[main/.style = {draw, circle, minimum size=2.5em}, scale=0.6, every node/.append style={transform shape}, 
        x1/.style={fill=dred, text=white},
		x2/.style={fill=dblue, text=white},
		x3/.style={fill=dgreen, text=white},
		x4/.style={fill=dorange, text=white}] 
		
\pgfmathsetmacro{\xsp}{1.8}
\pgfmathsetmacro{\ysp}{2.2}

\foreach \i in {1,...,4}
{
    \pgfmathsetmacro{\y}{-1*(\i - 1)*\ysp};
    \node[] (\i * 7) at (0,\y) {\large $\boldsymbol{x_{\i}}$};
    \foreach \j in {1,...,6}
    {
        \pgfmathtruncatemacro{\label}{\i*7 + \j};
        \pgfmathtruncatemacro{\prev}{(\label - 1)};
        \pgfmathsetmacro{\x}{\j * \xsp};
        \node[main,x\i] (\label) at (\x,\y) {\textbf{\j}};
        \ifthenelse{\j > 1}
        {\draw[-] (\label) -- (\prev);}
           
    }
    \pgfmathtruncatemacro{\ff}{(\i-1)*7 + 1};
    \pgfmathtruncatemacro{\lf}{(\i-1)*7 + 6};
    \pgfmathtruncatemacro{\fs}{\i*7 + 1};
    \pgfmathtruncatemacro{\ls}{\i*7 + 6};
    \ifthenelse{\i > 1}
    {
        \draw[->,>=stealth] (\ff.330) -- (\ls.150);
        \draw[->,>=stealth] (\lf.210) -- (\fs.30);
        \draw[->,>=stealth] (\ff.270) -- (\fs.90);
        \draw[->,>=stealth] (\lf.270) -- (\ls.90);
    }
}

\end{tikzpicture} 
\caption{Example $\Psi$ Connected Assignment Graph}
\label{fig:ham-cycle}
\end{figure}

Next, we need to ensure that Hamiltonian paths in $G$ encode \textbf{satisfying} truth assignments. In other words, we need to represent clauses in $G$. We do that by adding $k$ clause vertices corresponding to $k$ clauses. Each clause vertex $C_j$ should be connected to two vertices each sub-path belonging to its constituent variables, namely, $x_{i,2j-1}$ and $x_{i,2j}$. This results in the following properties:
\begin{enumerate}[label=\textbf{Property \arabic*:}, leftmargin=6em]
    \item No sub-path vertex is connected to more than one clause vertex. 
    \item There are two ways to visit $C_j$ in a Hamiltonian path, namely: $x_{i,2j-1} \rightarrow C_j \rightarrow x_{i,2j}$ and $x_{i,2j} \rightarrow C_j \rightarrow x_{i,2j-1}$.
\end{enumerate}
The two edges connecting $C_j$ to sub-path $x_i$ take the first direction (left to right) if clause $C_j$ contains $x_i$ unnegated, otherwise the edges are directed from right to left. In doing so, a variable assignment satisfies clause $C_j$ if its corresponding Hamiltonian path visits the $C_j$ vertex. The reason this is the case is that variable sub-paths can only be traversed in one direction in a Hamiltonian path, so if vertex $C_j$ is not oriented according to that direction, the truth assignment on that variable will not satisfy the clause. For example, as shown in Figure \ref{fig:ham-cycle2}, $C_1$ contains $x_1$ in positive form, so the edges connecting it to the $x_1$ sub-path are directed from left to right. Consequently, any Hamiltonian path that traverses $x_1$ from right to left (i.e., $x_1=$\textit{false}) will not be able to visit $C_1$ from $x_1$, and vice versa. 

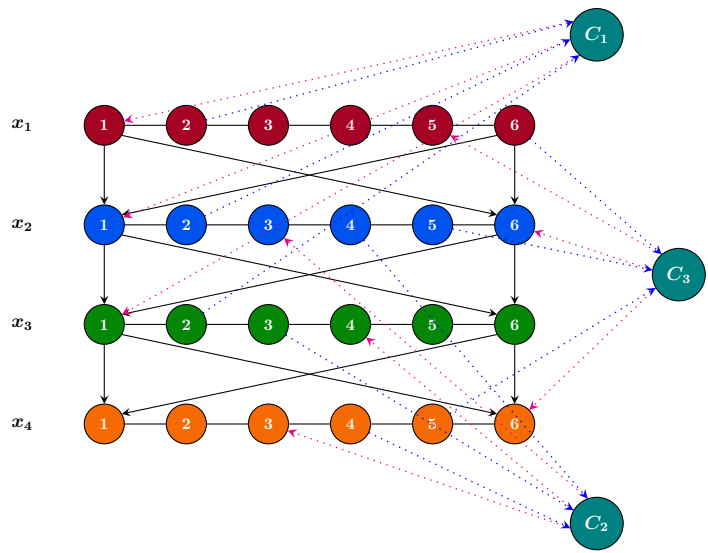
\begin{figure}[!h]
\centering

\begin{tikzpicture}[main/.style = {draw, circle, minimum size=2.5em}, scale=0.6, every node/.append style={transform shape}, 
        x1/.style={fill=dred, text=white},
		x2/.style={fill=dblue, text=white},
		x3/.style={fill=dgreen, text=white},
		x4/.style={fill=dorange, text=white},
		C1/.style={fill=teal, text=white, minimum size=3.3em},
		C2/.style={fill=teal, text=white, minimum size=3.3em},
		C3/.style={fill=teal, text=white, minimum size=3.3em}] 
		
\pgfmathsetmacro{\xsp}{1.8}
\pgfmathsetmacro{\ysp}{2.2}

\foreach \i in {1,...,4}
{
    \pgfmathsetmacro{\y}{-1*(\i - 1)*\ysp};
    \node[] (\i * 7) at (0,\y) {\large $\boldsymbol{x_{\i}}$};
    \foreach \j in {1,...,6}
    {
        \pgfmathtruncatemacro{\label}{(\i-1)*7 + \j};
        \pgfmathtruncatemacro{\prev}{(\label - 1)};
        \pgfmathsetmacro{\x}{\j * \xsp};
        \node[main,x\i] (\label) at (\x,\y) {\textbf{\j}};
        \ifthenelse{\j > 1}
        {\draw[-] (\label) -- (\prev);}
           
    }
    \pgfmathtruncatemacro{\ff}{(\i-2)*7 + 1};
    \pgfmathtruncatemacro{\lf}{(\i-2)*7 + 6};
    \pgfmathtruncatemacro{\fs}{(\i-1)*7 + 1};
    \pgfmathtruncatemacro{\ls}{(\i-1)*7 + 6};
    \ifthenelse{\i > 1}
    {
        \draw[->,>=stealth] (\ff.330) -- (\ls.150);
        \draw[->,>=stealth] (\lf.210) -- (\fs.30);
        \draw[->,>=stealth] (\ff.270) -- (\fs.90);
        \draw[->,>=stealth] (\lf.270) -- (\ls.90);
    }
}
\edef\tonode{{{1,8,15}, {10,18,24}, {5,13,27}}}
\edef\fromnode{{{2, 9, 16}, {11, 17, 25}, {6, 12, 26}}}
\edef\angst{{150, 110, 130}}
\node[main,C1] (C1) at (\xsp*7,2) {\large $\boldsymbol{C_1}$};
\node[main,C2] (C2) at (\xsp*7,-1*\ysp*4) {\large $\boldsymbol{C_2}$};
\node[main,C3] (C3) at (\xsp*8,-1*\ysp*1.5) {\large $\boldsymbol{C_3}$};
\foreach \i in {1,...,3}
{
    \foreach \z in {0,...,2}
    {
        \pgfmathsetmacro{\angt}{\angst[\i-1] + \z*40}
        \pgfmathsetmacro{\tox}{\tonode[\i-1][\z]}
        \pgfmathsetmacro{\fox}{\fromnode[\i-1][\z]}
        \draw[dotted,line width=0.5pt,->,>=stealth,magenta] (C\i.\angt) -- (\tox);
        \draw[dotted,line width=0.5pt,->,>=stealth, blue] (\fox) -- (C\i.\angt);
    }
    
};

\end{tikzpicture} 
\caption{Example $\Psi$ Clause-Aware Assignment Graph}
\label{fig:ham-cycle2}
\end{figure}

Now, in order to turn Hamiltonian paths into cycles, we add two vertices: a source vertex $s$, and a target vertex $t$ in the way shown in Figure \ref{fig:ham-cycle3}. That simplifies the structure of assignment Hamiltonian cycles by stipulating that they start and end at vertex $s$, without loss of generality. Note that it is very easy to find a cycle that visits all non-clause vertices as there are $2^n$ such cycles even if $\Psi$ itself is not satisfiable. The goal, however, is to be able to visit clause vertices as well, that is, our assignment path must satisfy (i.e., visit) all clause vertices. 

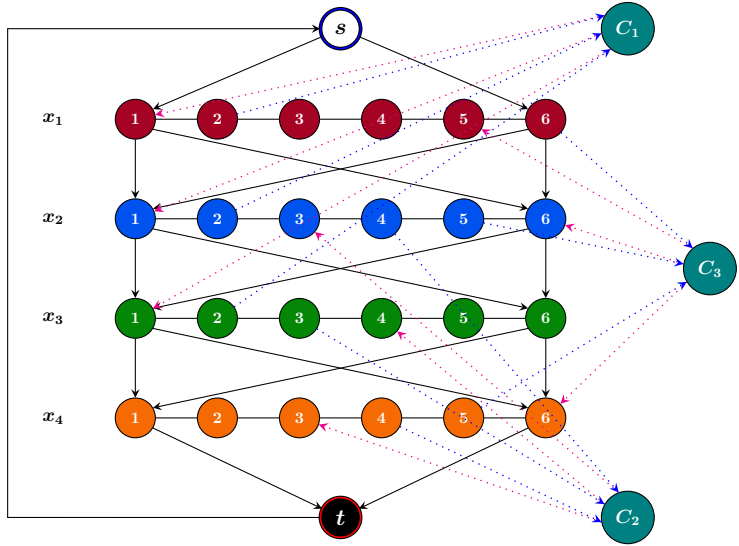
\begin{figure}[!h]
\centering

\begin{tikzpicture}[main/.style = {draw, circle, minimum size=2.5em}, scale=0.6, every node/.append style={transform shape}, 
        s/.style={double=blue, fill=white, text=black},
        t/.style={double=red, fill=black, text=white},
        x1/.style={fill=dred, text=white},
		x2/.style={fill=dblue, text=white},
		x3/.style={fill=dgreen, text=white},
		x4/.style={fill=dorange, text=white},
		C1/.style={fill=teal, text=white, minimum size=3.3em},
		C2/.style={fill=teal, text=white, minimum size=3.3em},
		C3/.style={fill=teal, text=white, minimum size=3.3em}] 
		
\pgfmathsetmacro{\xsp}{1.8}
\pgfmathsetmacro{\ysp}{2.2}

\foreach \i in {1,...,4}
{
    \pgfmathsetmacro{\y}{-1*(\i - 1)*\ysp};
    \node[] (\i * 7) at (0,\y) {\large $\boldsymbol{x_{\i}}$};
    \foreach \j in {1,...,6}
    {
        \pgfmathtruncatemacro{\label}{(\i-1)*7 + \j};
        \pgfmathtruncatemacro{\prev}{(\label - 1)};
        \pgfmathsetmacro{\x}{\j * \xsp};
        \node[main,x\i] (\label) at (\x,\y) {\textbf{\j}};
        \ifthenelse{\j > 1}
        {\draw[-] (\label) -- (\prev);}
           
    }
    \pgfmathtruncatemacro{\ff}{(\i-2)*7 + 1};
    \pgfmathtruncatemacro{\lf}{(\i-2)*7 + 6};
    \pgfmathtruncatemacro{\fs}{(\i-1)*7 + 1};
    \pgfmathtruncatemacro{\ls}{(\i-1)*7 + 6};
    \ifthenelse{\i > 1}
    {
        \draw[->,>=stealth] (\ff.330) -- (\ls.150);
        \draw[->,>=stealth] (\lf.210) -- (\fs.30);
        \draw[->,>=stealth] (\ff.270) -- (\fs.90);
        \draw[->,>=stealth] (\lf.270) -- (\ls.90);
    }
}
\edef\tonode{{{1,8,15}, {10,18,24}, {5,13,27}}}
\edef\fromnode{{{2, 9, 16}, {11, 17, 25}, {6, 12, 26}}}
\edef\angst{{150, 110, 130}}
\node[main,C1] (C1) at (\xsp*7,2) {\large $\boldsymbol{C_1}$};
\node[main,C2] (C2) at (\xsp*7,-1*\ysp*4) {\large $\boldsymbol{C_2}$};
\node[main,C3] (C3) at (\xsp*8,-1*\ysp*1.5) {\large $\boldsymbol{C_3}$};
\foreach \i in {1,...,3}
{
    \foreach \z in {0,...,2}
    {
        \pgfmathsetmacro{\angt}{\angst[\i-1] + \z*40}
        \pgfmathsetmacro{\tox}{\tonode[\i-1][\z]}
        \pgfmathsetmacro{\fox}{\fromnode[\i-1][\z]}
        \draw[dotted,line width=0.5pt,->,>=stealth,magenta] (C\i.\angt) -- (\tox);
        \draw[dotted,line width=0.5pt,->,>=stealth, blue] (\fox) -- (C\i.\angt);
    }
    
};

\node[main,s] (s) at (\xsp*3.5,2) {\Large $\boldsymbol{s}$};
\draw[->,>=stealth] (s) -- (1.30);
\draw[->,>=stealth] (s) -- (6.150);

\node[main,t] (t) at (\xsp*3.5,-1*\ysp*4) {\Large $\boldsymbol{t}$};
\draw[->,>=stealth] (22.-30) -- (t);
\draw[->,>=stealth] (27.-150) -- (t);

\draw[->,>=stealth] (t) -- (-1, -1*\ysp*4) -- (-1,2) -- (s);

\end{tikzpicture} 
\caption{Example $\Psi$ Final Assignment Graph}
\label{fig:ham-cycle3}
\end{figure}

To finalize the correctness of this reduction, we verify both directions of the correspondence between satisfying assignments on $\Psi$ and Hamiltonian cycles on $G$:

$(\Rightarrow):$ If $\Psi$ is satisfiable, then there must exist a satisfying assignment $f: \{x_1,...,x_n\} \rightarrow \{\text{\textit{true}, \textit{false}}\}$. We can then construct a Hamiltonian cycle on $G$ as follows:
\begin{itemize}
    \item Start at vertex $s$.
    \item For every variable $x_i$, traverse its sub-path in the direction that matches its truth value while visiting any reachable unvisited clause vertices along the way.
    \item Go to vertex $t$ and return to $s$.
\end{itemize}

$(\Leftarrow):$ If $G$ contains a Hamiltonian cycle $\mathcal{H}$, then we can construct a satisfying assignment $f: \{x_1,...,x_n\} \rightarrow \{\text{\textit{true}, False}\}$ for $\Psi$ as follows.

\[ f(x_i) = \begin{cases} 
      \text{\textit{true}}, & (x_{i,1},...,x_{i,2k})\in \mathcal{H} \\
      \text{\textit{false}}, & (x_{i,2k},...,x_{i,1})\in \mathcal{H}
   \end{cases}
\]

This assignment is guaranteed to satisfy all clauses since $\mathcal{H}$ must visit all clause vertices. Thus, this reduction is indeed correct.

Finally, it remains to show that this reduction is polynomial-time. This is verily the case since $|G|=2n|\Psi| + |\Psi|+2= (2n+1)|\Psi| + 2 = (2n+1)k + 2$, hence the size of $G$ must also be a polynomial. Therefore, constructing this graph takes polynomial-time.

\end{proof}

\clearpage

\section{3-SAT to 3-COLOR}
Graph coloring is one of most important concepts of graph theory. It essentially involves assigning each vertex in a graph with some color to satisfy certain properties or constraints. For $k$-coloring problems, we ask whether we can color a graph with at most $k$ colors such that no two adjacent vertices have the same color. 

\begin{definition}
A $k$-coloring of an undirected graph $G$ is a function $f: V(G) \rightarrow \{1, ..., k\}$ such that $\{u,v\}\in E(G) \rightarrow f(u) \neq f(v)$.

\begin{remark}
We say graph $G$ is $k$-colorable if such function exists for it.
\end{remark}
\end{definition}

Based on this definition, testing 2-colorability is equivalent to performing a bipartite check which can be done in linear time. Simply speaking, we can just traverse graph vertices in breadth-first order, coloring the neighborhood for every vertex with an opposite color until we hit a conflict of colors. 

\begin{algorithm}[H]
\SetAlgoLined
\SetKwInOut{Input}{Input}
\SetKwRepeat{Do}{do}{while}
\Input{Undirected graph $G$}
\KwResult{YES if $G$ is 2-colorable, otherwise, NO}
Let $\mathcal{BFS}(G) := \text{the vertex breadth-first traversal order for } G$ \\
Let $\mathcal{C}(u) := \text{color of vertex } u$ \\
Initialize $\mathcal{C}(u) = -1, \forall u \in V(G)$

\ForEach{vertex $u \in \mathcal{BFS}(G)$}{
    \uIf{$\mathcal{C}(u) = -1$}{%
            $\mathcal{C}(u) := 0$
    }
    \ForEach{vertex $v \in N(u)$}{
    
        \uIf{$\mathcal{C}(v) = -1$}{%
            $\mathcal{C}(v) := 1 - \mathcal{C}(u)$
        }
        \uElseIf{$\mathcal{C}(v) \neq \mathcal{C}(u)$}{
            \Return NO
        }
    }
}
\Return YES \\

 \caption{Bipartite Check (2-COLOR)}
\end{algorithm}

As such, 2-COLOR $\in P$. Sadly, much like the case of 2-SAT and 3-SAT, 3-COLOR is NP-complete despite it seeming like a slightly harder version of 2-COLOR. Firstly, it is quite easy to see why 3-COLOR $\in NP$ since its certificates are vertex color assignments which are the same size of the graph and can be checked in polynomial time by traversing edges, checking if two end-vertices have the same color. Secondly, we shall now show that 3-SAT reduces to 3-COLOR thereby proving it intractable.

\begin{theorem}
3-SAT $\leq_{\mathcal{P}}$ 3-COLOR.
\end{theorem}

\begin{proof}
To establish the reduction, for any 3-SAT Boolean formula $\Psi$ in variables $\{x_1,...,x_n\}$, we construct a graph $G$ such that $G$ is 3-colorable if and only if $\Psi$ is satisfiable. We start by adding two vertices for every variable in $\Psi$: one for $x_i$, and another for $\neg x_i$. We now need to ensure a coloring on the vertices of $G$ corresponds to a valid assignment. To achieve this, we add a triangle subgraph with three vertices T (True), F (False), and B (Base), and connect every variable and its negation with B as shown below.

\begin{figure}[!h]
\centering

\begin{tikzpicture}[main/.style = {draw, circle, minimum size=3.3em}, scale=0.7, every node/.append style={transform shape}, 
        B/.style={fill=cyan, text=white},
		T/.style={fill=darkgreen, text=white},
		F/.style={fill=red, text=white}] 
\node[main,B] (1) at (0,0) {\Large \textbf{B}}; 
\node[main,F] (2) at (1.5,-1.5) {\Large \textbf{F}};
\node[main,T] (3) at (-1.5,-1.5) {\Large \textbf{T}}; 

\node[main] (4) at (-4.9,1) {\Large $x_1$};
\node[main] (5) at (-3.9,2.4) {\Large $\neg x_1$}; 

\node[main] (6) at (-0.9,3.5) {\Large $x_2$};
\node[main] (7) at (0.9,3.5) {\Large $\neg x_2$}; 

\node[main] (8) at (4.9,1) {\Large $x_3$}; 
\node[main] (9) at (3.9,2.4) {\Large $\neg x_3$};

\draw[-] (1) -- (2);
\draw[-] (2) -- (3);
\draw[-] (3) -- (1);

\draw[-] (4) -- (5);
\draw[-] (4) -- (1);
\draw[-] (5) -- (1);

\draw[-] (6) -- (7);
\draw[-] (6) -- (1);
\draw[-] (7) -- (1);

\draw[-] (8) -- (9);
\draw[-] (8) -- (1);
\draw[-] (9) -- (1);

\end{tikzpicture} 
\caption{Subgraph of $G$}
\label{fig:3color}
\end{figure}
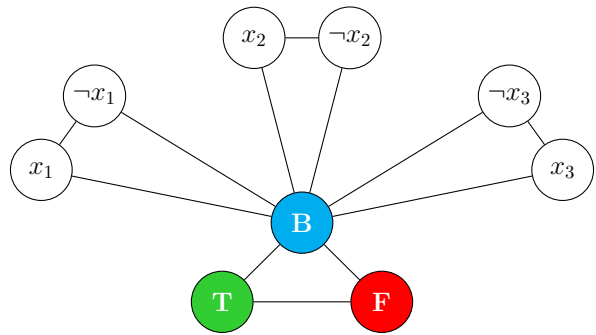

These connections ensure the following properties:
\begin{enumerate}[label=\textbf{Property \arabic*:}, leftmargin=8em]
    \item Each variable gets a different color from its negation.
    \item T, F, and B vertices get different colors.
    \item Any 3-coloring of the graph will map to a valid truth assignment since no literal can take the neutral base color, so they will either be green (i.e., True) or red (i.e., False).
\end{enumerate}

Next, we need to represent clauses in $G$. To emulate a clause structure (i.e., disjunction), we need to add a few extra vertices and connections between these vertices and the original literal vertices we already have. This extra gadget is shown in Figure \ref{fig:3color_1}. Note that $t_1,t_2,$ and $t_3$ are literal vertices (i.e., could either be $x_i$ or $\neg x_i$) of one clause, so this gadget should be added for every clause in $\Psi$.

\begin{figure}[!h]
\centering

\begin{tikzpicture}[main/.style = {draw, circle, minimum size=2.8em}, scale=0.6, every node/.append style={transform shape}, 
        B/.style={fill=cyan, text=white},
		T/.style={fill=darkgreen, text=white},
		F/.style={fill=red, text=white},
		D/.style={double, double distance=2pt, outer sep=2pt}] 
		
\node[main] (1) at (0,0) {};

\node[main] (2) at (4,2) {};
\node[main] (3) at (-4,2) {};
\node[main] (4) at (4,-2) {};
\node[main] (5) at (-4,-2) {};

\node[main] (6) at (0,3.5) {};
\node[main,T] (7) at (0,-3.5) {\Large \textbf{T}};

\node[main,F] (8) at (7,2) {\Large \textbf{F}};
\node[main,D] (9) at (-4,-5) {\Large $t_1$};

\node[main,D] (10) at (4,-5) {\Large $t_2$};
\node[main,D] (11) at (2,-5) {\Large $t_3$};

\draw[-] (1) -- (6);
\draw[-] (1) -- (7);
\draw[-] (1) -- (11);

\draw[-] (7) -- (4);
\draw[-] (7) -- (5);
\draw[-] (7) -- (3);

\draw[-] (4) -- (2);
\draw[-] (4) -- (10);

\draw[-] (8) -- (2);
\draw[-] (9) -- (5);

\draw[-] (3) -- (6);
\draw[-] (3) -- (5);

\draw[-] (2) -- (6);

\end{tikzpicture} 
\caption{Gadget subgraph $\mathcal{H}$ of clause $(t_1\lor t_2\lor t_3)$}
\label{fig:3color_1}
\end{figure}
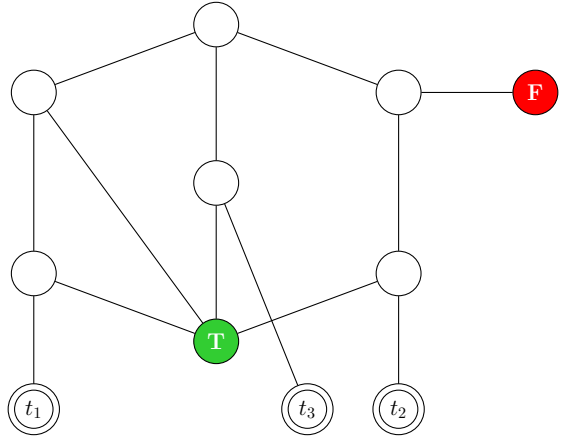

Note that the property we need to ensure is that $\mathcal{H}$ is 3-colorable if and only if clause $(t_1\lor t_2\lor t_3)$ is True, that is, at least one of $t_1,t_2,$ and $t_3$ is True. We can demonstrate that this is indeed the case by first showing the graph coloring that results from having all literals set to False:

\begin{figure}[!h]
\centering

\begin{tikzpicture}[main/.style = {draw, circle, minimum size=2.8em}, scale=0.6, every node/.append style={transform shape}, 
        B/.style={fill=cyan, text=white},
		T/.style={fill=darkgreen, text=white},
		F/.style={fill=red, text=white},
		D/.style={double, double distance=2pt, outer sep=2pt}] 
		
\node[main,B] (1) at (0,0) {};

\node[main,T] (2) at (4,2) {};
\node[main,F] (3) at (-4,2) {};
\node[main,B] (4) at (4,-2) {};
\node[main,B] (5) at (-4,-2) {};

\node[main] (6) at (0,3.5) {???};
\node[main,T] (7) at (0,-3.5) {\Large \textbf{T}};

\node[main,F] (8) at (7,2) {\Large \textbf{F}};
\node[main,D,F] (9) at (-4,-5) {\Large $t_1$};

\node[main,D,F] (10) at (4,-5) {\Large $t_2$};
\node[main,D,F] (11) at (2,-5) {\Large $t_3$};

\draw[-] (1) -- (6);
\draw[-] (1) -- (7);
\draw[-] (1) -- (11);

\draw[-] (7) -- (4);
\draw[-] (7) -- (5);
\draw[-] (7) -- (3);

\draw[-] (4) -- (2);
\draw[-] (4) -- (10);

\draw[-] (8) -- (2);
\draw[-] (9) -- (5);

\draw[-] (3) -- (6);
\draw[-] (3) -- (5);

\draw[-] (2) -- (6);

\end{tikzpicture} 
\caption{Gadget $\mathcal{H}$ Coloring When All Clause Variables Are False}
\label{fig:3color_2}
\end{figure}
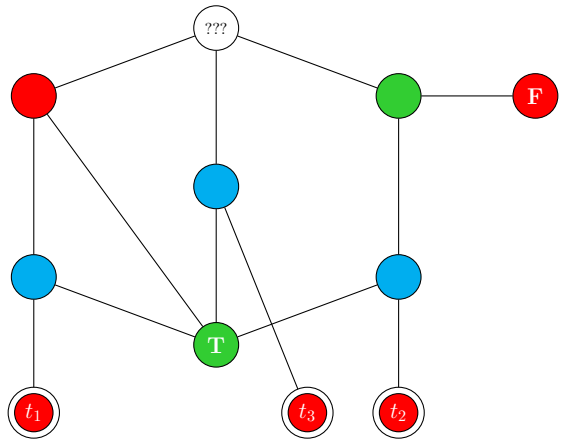

We can see in Figure \ref{fig:3color_2} that the top vertex cannot take any of the three colors, making gadget $\mathcal{H}$ not 3-colorable when its clause is False. This essentially proves the forward direction. To prove the reverse direction, we let one of the three literals be True and show that $\mathcal{H}$ would be 3-colorable. Note that it makes no difference which literal we set to True since disjunction is commutative. 

\begin{figure}[!h]
\centering

\begin{tikzpicture}[main/.style = {draw, circle, minimum size=2.8em}, scale=0.6, every node/.append style={transform shape}, 
        B/.style={fill=cyan, text=white},
		T/.style={fill=darkgreen, text=white},
		F/.style={fill=red, text=white},
		D/.style={double, double distance=2pt, outer sep=2pt}] 
		
\node[main,B] (1) at (0,0) {};

\node[main,T] (2) at (4,2) {};
\node[main,B] (3) at (-4,2) {};
\node[main,B] (4) at (4,-2) {};
\node[main,F] (5) at (-4,-2) {};

\node[main,F] (6) at (0,3.5) {};
\node[main,T] (7) at (0,-3.5) {\Large \textbf{T}};

\node[main,F] (8) at (7,2) {\Large \textbf{F}};
\node[main,D,T] (9) at (-4,-5) {\Large $t_1$};

\node[main,D,F] (10) at (4,-5) {\Large $t_2$};
\node[main,D,F] (11) at (2,-5) {\Large $t_3$};

\draw[-] (1) -- (6);
\draw[-] (1) -- (7);
\draw[-] (1) -- (11);

\draw[-] (7) -- (4);
\draw[-] (7) -- (5);
\draw[-] (7) -- (3);

\draw[-] (4) -- (2);
\draw[-] (4) -- (10);

\draw[-] (8) -- (2);
\draw[-] (9) -- (5);

\draw[-] (3) -- (6);
\draw[-] (3) -- (5);

\draw[-] (2) -- (6);

\end{tikzpicture} 
\caption{Gadget $\mathcal{H}$ 3-Coloring When $t_1=$ True}
\label{fig:3color_3}
\end{figure}
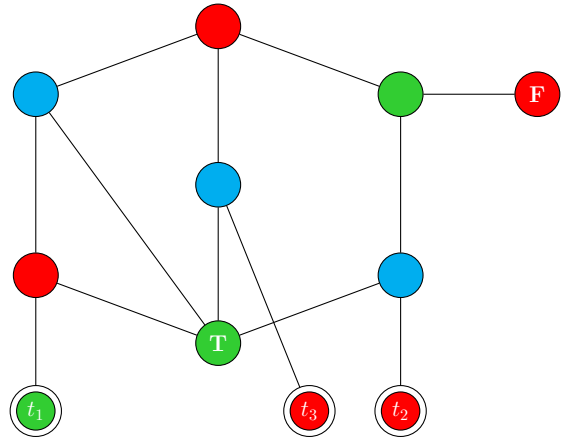

As shown in Figure \ref{fig:3color_3}, $\mathcal{H}$ is indeed 3-colorable, and the same scheme would apply if we make any of the two other literals True (i.e., green) as well. Thus, 3-colorability of a gadget subgraph corresponds to the satisfiability of its respective clause. 

The same applies to the whole graph $G$ with all clauses mapped out to gadget subgraphs since 3-colorings of $G$ correspond to valid truth assignments (property 3) that satisfies all clauses, hence satisfying $\Psi$. Thus, the reduction is correct.

Finally, it remains to show that this reduction is polynomial. That is indeed the case because $|G|=2n + 6  |\Psi|$ where $n$ is the number of variables in $\Psi$ and $|\Psi|$ is the number of its clauses. Therefore, constructing this graph takes polynomial-time.

\end{proof}



\bibliographystyle{plain}
\bibliography{ref.bib}

\end{document}